\documentclass[times,twocolumn,final,5p]{elsarticle}

\usepackage{framed,multirow}

\usepackage{amssymb}
\usepackage{latexsym}

\usepackage{url}
\usepackage{color,colortbl,booktabs}
\usepackage[table,x11names,dvipsnames,table]{xcolor}
\definecolor{lightgray}{gray}{0.95}

\usepackage{hyperref}
\hypersetup{
    colorlinks=true,
    linkcolor=blue,
}

\usepackage{float,tablefootnote,threeparttable,wrapfig}

\usepackage{booktabs}
\usepackage{subcaption}

\usepackage{lipsum}
\usepackage{color}
\usepackage[ruled,vlined]{algorithm2e}
\usepackage{diagbox} 
\usepackage{tabu}
\usepackage{arydshln}

\newcommand{\revision}[1]{\textcolor{black}{#1}} 
\newcommand{\revisionSecond}[1]{\textcolor{black}{#1}}

\usepackage{xcolor}
\definecolor{ForestGreen}{rgb}{0.13, 0.55, 0.13}
\usepackage{pifont}
\usepackage{amsmath} 
\usepackage{amssymb}  
\usepackage{diagbox} 
\usepackage{tabu}
\usepackage{algpseudocode}
\usepackage{epsfig}
\usepackage{hyperref}
\usepackage{booktabs}  
\usepackage{multirow}  
\usepackage{array}

\usepackage{overpic}

\journal{Medical Image Analysis}

\begin{document}

\begin{frontmatter}

\title{Speckle2Self: Self-Supervised Ultrasound Speckle Reduction \\Without Clean Data}

\author[1,2]{Xuesong Li}
\author[1,2]{Nassir Navab}
\author[1,2]{Zhongliang Jiang\corref{cor1}}

\cortext[cor1]{
Corresponding author at: Technical University of Munich, Faculty of informatics -- I16, Boltzmannstr. 3, 85748 Garching bei M{\"u}nchen. \\
This work involved human subjects in its research. Approval of all ethical and experimental procedures and protocols was granted by Institutional Review Board, No. 2022-87-S-KK, Declaration of Helsinki.
}
\ead{zl.jiang@tum.de}

\address[1]{Computer Aided Medical Procedures, Technical University of Munich, Munich, Germany}
\address[2]{Munich Center for Machine Learning (MCML), Munich, Germany}

\begin{abstract} 
Image denoising is a fundamental task in computer vision, particularly in medical ultrasound (US) imaging, where speckle noise significantly degrades image quality. Although recent advancements in deep neural networks have led to substantial improvements in denoising for natural images, these methods cannot be directly applied to US speckle noise, as it is not purely random. Instead, US speckle arises from complex wave interference within the body microstructure, making it tissue-dependent. This dependency means that obtaining two independent noisy observations of the same scene, as required by pioneering Noise2Noise, is not feasible. Additionally, blind-spot networks also cannot handle US speckle noise due to its high spatial dependency. To address this challenge, we introduce \revision{Speckle2Self}, a novel self-supervised algorithm for speckle reduction using only single noisy observations. The key insight is that applying a multi-scale perturbation (MSP) operation introduces tissue-dependent variations in the speckle pattern across different scales, while preserving the shared anatomical structure. This enables effective speckle suppression by modeling the clean image as a low-rank signal and isolating the sparse noise component.
To demonstrate its effectiveness, Speckle2Self is comprehensively compared with conventional filter-based denoising algorithms and SOTA learning-based methods, using both realistic simulated US images and human carotid US images. Additionally, data from multiple US machines are employed to evaluate model generalization and adaptability to images from unseen domains. Project page: \url{https://noseefood.github.io/us-speckle2self/}
\end{abstract}


\end{frontmatter}
    

\section{Introduction}
\label{introduction}


\begin{figure}[ht!]
    \centering
    \includegraphics[width=0.85\linewidth]{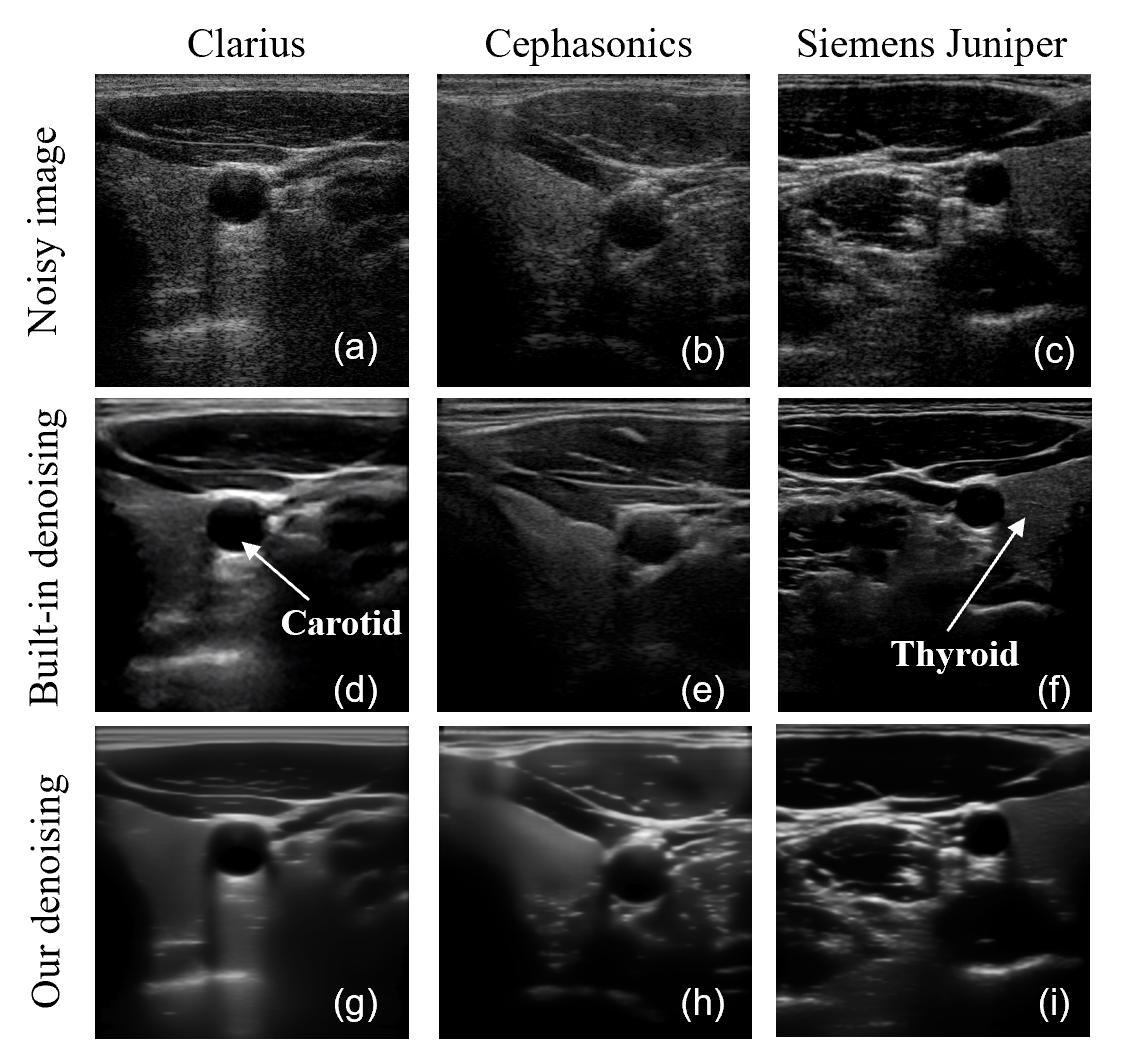}
    \caption{Top row demonstrates the noisy US image of a human carotid from three different commercial US machines. The middle row shows the results obtained by using their default setting for a carotid. The bottom row depicts the denoising results obtained using the proposed self-supervised Speckle2Self algorithm. \revision{Only the noisy and denoised images using built-in filters acquired from the Clarius device were acquired simultaneously. In contrast, these pairs from the Cephasonics and Juniper machines were captured at different time points from the same location, which may introduce slight differences due to physiological motion and human tremors.}}
    \label{fig:speckle_noise_illus}
\end{figure}

\par
Medical ultrasound (US) is one of the most important imaging modalities in modern clinical practices due to its affordability, non-invasiveness and real-time capabilities~\cite{jiang2023robotic,bi2024machine}. US imaging visualises internal anatomical structures by emitting high-frequency acoustic waves (typically 2$-$15 MHz) into the body and detecting echoes scattered from tissue interfaces~\cite{szabo2013diagnostic}. Compared to Computed Tomography (CT) and Magnetic Resonance Imaging (MRI), US images generally suffer from lower image quality~\cite{kang2024deblurring, stevens2024dehazing,calis2025speckle, mwikirize2018signal}, primarily due to speckle noise—one of the most prominent artefacts in B-mode imaging.
This speckle noise arises from the coherent summation of echoes scattered by small-scale tissue structures (e.g., cells) and manifests as grainy patterns that degrade image clarity and contrast~\cite{krissian2005speckle}.
The degradation is especially pronounced in homogeneous regions, such as the thyroid (see Fig.~\ref{fig:speckle_noise_illus} (a–c)), where microstructures are evenly distributed. 
Beyond visual degradation, speckle noise also impairs the accurate identification of both normal and pathological tissues, thereby reducing the effectiveness of computer-aided diagnosis~\cite{huang2018machine} and downstream tasks like segmentation~\cite{lee2022speckle,bi2023mi,huang2025vibnet,huang2023motion,jiang2024intelligent},  registration~\cite{ramalhinho2020registration, song2021cross,jiang2024class} and anomaly detection in US images~\cite{bi2025synomaly,zhou2025ultraad}. 
To enhance diagnostic quality, modern US systems typically combine advanced hardware with tailored temporal/spatial filtering techniques to mitigate speckle. Fig.~\ref{fig:speckle_noise_illus}~(d-f) illustrates representative post-processed B-mode images acquired from three commercial US Systems—Clarius (Clarius Mobile Health, Canada), Cephasonics (Cephasonics Ultrasound, USA), and Juniper (Siemens Healthineers, Germany)—using their default manufacturer-recommended settings to image nearly the same anatomical region.

However, the high-performance optimised US devices may not be accessible, particularly in rural or resource-limited healthcare settings. Recent advancements in point-of-care ultrasound (POCUS) emphasizes the need for a universal speckle denoising method that can be applied across different machines, especially for portable and low-cost devices. In this context, the ideal solution should be lightweight and data-efficient, without requiring additional signals such as low-level raw frequency data~\cite{sharifzadeh2024mitigating}.



\subsection{Conventional Denoising Approaches}
\par

Speckle noise in US is highly tissue-dependent and inherently approximately multiplicative, meaning that its intensity can vary significantly—often with fluctuations comparable to those of the tissue signal itself~\cite{mei2019phase, wagner1983statistics}. In addition, unlike pixel-wise Gaussian noise, speckle noise typically appears in clusters, and the grainy pattern varies in intensity across different tissues, which blend with image features (see Fig.~\ref{fig:fig_nosie_comparision}). These characteristics make it challenging to remove speckle while preserving essential tissue features.

To tackle this challenge, a set of despeckling filters, including local adaptive filters~\cite{frost1982model, zhang2015wavelet, tay2010ultrasound}, non-local means (NLM) filters~\cite{coupe2009nonlocal, zhou2019iterative, zhu2017non, xie2016multispectral}, and diffusion filters~\cite{perona1990scale, yu2002speckle, aja2006estimation, krissian2007oriented, zhou2014doubly}, were presented. The core idea of local adaptive filters is to remove speckle in a small patch by identifying nearby patches with similar local features and applying a weighted averaging scheme in different space. This approach, however, often struggles to preserve sharp boundaries~\cite{chen2010ramp} and may introduce additional artefacts during the filtering process.
In addition, such local filters also rely heavily on the selection of parameters like local window size and shape. 
One of the most successful adaptations for US images, Optimal Bayesian Non-Local Means (OBNLM)~\cite{coupe2009nonlocal}, was specifically developed to address speckle noise. 
However, OBNLM and its variants~\cite{zhu2017non, zhou2019iterative} often result in incomplete speckle suppression and are also prone to introducing artifacts.
In terms of diffusion filtering, the well-known anisotropic diffusion (AD) approach~\cite{perona1990scale} has been adapted for speckle reduction in US imaging. Speckle-Reducing AD (SRAD)~\cite{yu2002speckle} incorporated noise-related parameters into the diffusion coefficient to better target speckle noise, while Detail-Preserving AD (DPAD)~\cite{aja2006estimation} further refines this approach by incorporating a noise estimator, enhancing despeckling performance while maintaining structural details. These methods are non-learning-based without the need for clean ground truth images or explicit noise models.

\subsection{Fully-Supervised Denoising Approaches}
\par
Recent advances in deep neural networks have generated significant interest in bypassing traditional statistical modeling of signal corruption by instead learning to map corrupted observations directly to clean images. Convolutional neural networks (CNNs) have been employed to establish regression models between paired corrupted and clean images. 
DnCNN~\cite{zhang2017beyond} is a representative example, achieving superior performance over classical denoising techniques. Building on this foundation, a variety of denoising networks~\cite{gu2019self, guo2019toward, lefkimmiatis2018universal, mao2016image, plotz2018neural, tai2017memnet} have been proposed to further improve performance.

However, even for natural images in computer vision, acquiring clean ground-truth data in real-world scenarios is often impractical. This challenge becomes more pronounced in the field of medical imaging, where clean references are generally unavailable. 
In particular, US imaging poses a unique difficulty: unlike high-resolution modalities such as CT, where pseudo-clean images can be acquired through high-dose protocols, US data is inherently degraded by tissue-dependent, deterministic speckle, making it impossible to obtain a truly clean or even approximated ground-truth image.

\subsection{Self-Supervised Denoising Approaches}
\label{sec:speckle}
To overcome this limitation, a range of self-supervised methods have been proposed and recently gained prominence in the field. Three popular approaches have emerged. \revision{The first one is Deep Image Prior (DIP), which} uses a neural network to learn a denoising prior directly from a single noisy image without external training data~\cite{ulyanov2018deep}. Although DIP has demonstrated effectiveness across various applications, it requires image-specific optimisation, making it computationally intensive and unsuitable for real-time applications. 

The second category involves training a denoising network using two independent noisy observations of the same scene, as proposed in Noise2Noise (N2N)~\cite{lehtinen2018noise2noise}. However, acquiring such independent observations is highly impractical in US imaging, where speckle noise is inherently deterministic. Unlike natural image noise, which often originates from random sensor fluctuations, US speckle patterns remain almost consistent in repeated observations of the same scene with identical configurations, violating the independence assumption required by N2N. \revisionSecond{Recently,~\cite{asgariandehkordi2025lightweight,jung2024unsupervised,cho2024deep} follow the design similar to N2N by obtaining two noisy observations of the same scene using the advanced technology of plane wave imaging, and have demonstrated promising performance in ultrasound image denoising. However, this technology has not yet been widely integrated into the commercial US device, particularly in the low-cost portable setting.}


In addition, the third class of methods focuses on cases where only single noisy images are available. These approaches aim to extract multiple subsets with approximately independent noise distributions from each noisy image, enabling the use of N2N-style training without requiring clean references. The most well-known method in this category is Noise2Void (N2V)~\cite{krull2019noise2void}, which is based on blind-spot networks (BSN). It does not require multiple observations of the same scene and has shown promising performance. However, it strongly relies on the assumption that noise is spatially independent within a single image—an assumption that often does not hold in real-world scenarios. To address this, subsequent improvements such as AP-BSN~\cite{lee2022ap} and C-BSN~\cite{jang2023self} have introduced new strategies to reduce spatial correlation in real-world noise, aiming to improve the robustness of BSN-based methods. 
Beyond BSN-based approaches, other methods such as Noise2Self~\cite{batson2019noise2self} and Neighbor2Neighbor~\cite{huang2021neighbor2neighbor} adopt different schemes to generate sub-images from single noisy inputs, each assumed to contain independent noise realisations, enabling self-supervised training under similar principles.

However, in US imaging, speckle exhibits strong spatial and tissue-dependent correlations. Unlike pixel-level random noise, speckle appears as large-scale, structured patterns, making it difficult to generate sub-images with uncorrelated speckle noise. As a result, these methods based on single noisy images, including BSN-based approaches, are generally unsuitable for speckle reduction in US imaging. 

In addition, there exists a distinct category of methods that aim to enhance performance by incorporating explicit noise modelling, such as~\cite{laine2019high, wu2020unpaired}. However, these noise model assumptions are often effective only in synthetic experiments where the noise distribution is known, and performance tends to degrade significantly on real-world noisy images where the noise characteristics are unknown~\cite{huang2021neighbor2neighbor}. In US imaging, it is impossible to obtain a speckle-free image, nor is there a universal statistical model for US speckle. Furthermore, unlike photography, where the raw file can provide the noise model for each image, no equivalent noise model exists in US imaging.
\par

\subsection{Proposed Approach}
\par
In this work, we propose \revision{Speckle2Self}, a new self-supervised despeckling framework tailored for US images. Since it is challenging to obtain two independent noisy observations of the same scene in US imaging, Noise2Noise and its variants cannot be directly applied to remove the tissue-dependent speckle. \revision{To address this, Speckle2Self applies a multi-scale perturbation (MSP) operation to a single noisy image, introducing discriminant variations in speckle patterns while keeping the underlying anatomical structure unchanged. This allows the model to learn to separate consistent structural information from inconsistent speckle noise.}
The main contributions of this work are summarised as follows:
\begin{itemize}
    \item We introduce \revision{a novel self-supervised despeckling framework specifically designed to address the challenge of tissue-dependent US speckle, using only single noisy images.} By eliminating the need for noisy-clean pairs or multiple observations of the same scene, our approach is uniquely suited for real-world US imaging scenarios where such clean data is typically unavailable. 
    
    \item We first leverage the MSP operation to create distinctive variations in US speckle patterns while preserving essential anatomical structures, establishing a robust foundation for effective speckle reduction using CNNs.
    
    \item We conducted comprehensive experiments on both simulated US images and in-vivo carotid images, and the proposed method significantly outperforms conventional US speckle reduction methods and popular learning-based denoising methods. In addition, the cross-device validation demonstrates the robustness and generalisation capabilities of the proposed method on unseen carotid images acquired from two additional US systems, highlighting its adaptability to diverse setups. 
\end{itemize}

\begin{figure}[ht!]
\centering
    \includegraphics[width=0.80\linewidth]{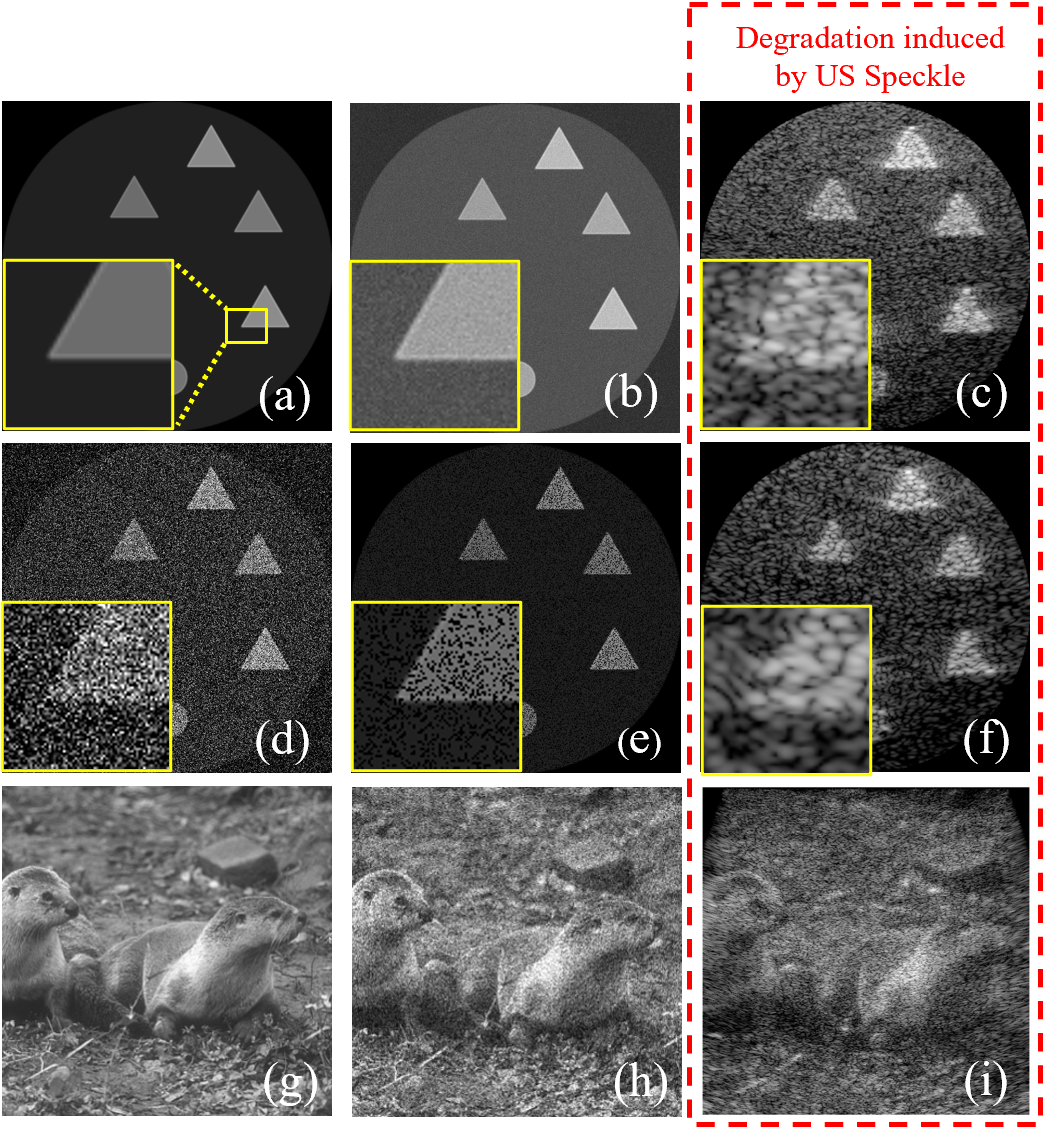}
    \caption{Intuitive visualization of the image quality degradation caused by different types of noise. The first two rows present visual comparisons based on synthetic data: (a) is the clean target; (c) and (f) are the synthetic US images generated using a relatively low and high US signal central frequency (5.63 MHz vs. 3.75 MHz) in the physic-based Matlab Ultrasound Toolbox~\cite{garcia2022simus, cigier2022simus, perrot2021so}, respectively; (b), (d) and (e) are the synthetic images by adding Gaussian($\sigma=75$), Salt-and-Pepper($p=0.25$) and Poisson noise($\lambda=50$) to the clean target, respectively. The third row shows visual comparisons based on natural images from the BSD68 dataset~\cite{MartinFTM01}: (g) is the clean target, (h) is the image corrupted with Gaussian noise at ($\sigma=75$), and (i) is corresponding simulated US image using central frequency at 7.50 MHz. \revision{All images are grayscale with intensity values in the range [0, 255] and have a resolution of 512×512 pixels\revisionSecond{, and correspond to a physical size of 45 mm × 50 mm (width × depth)}.} It is evident that due to the limited spatial resolution of US imaging and the presence of speckles, fine details in the natural images after US simulation are largely lost.}
    \label{fig:fig_nosie_comparision}
\end{figure}


\section{Preliminaries}
\label{sec:preliminary}
This section provides an overview of the \revision{US} speckle formation model, followed by a brief review of the fully supervised Noise2Clean (N2C) paradigm and popular self-supervised denoising approaches, including \revision{N2N}~\cite{lehtinen2018noise2noise} and \revision{N2V}~\cite{krull2019noise2void}.

\subsection{Ultrasound Image Formation}
\label{sec:us_property}
\par
In medical ultrasound imaging, speckle arises from the coherent interference of echoes scattered by small-scale tissue structures. It is typically modelled as multiplicative noise, where the observed signal is the product of a clean tissue signal and a speckle component. At the raw radio-frequency (RF) signal level, this can be expressed as:
\begin{equation}~\label{eq:speckle_noise}
  \mathbf{Y}_{(RF)}  = \mathbf{X}_{(RF)}\odot \mathbf{N}_{(RF)}
\end{equation}
\revision{where $\mathbf{Y}_{(RF)}$ represents the observed degraded RF signal, $\mathbf{X}_{(RF)}$ is the clean underlying signal, and $\mathbf{N}_{(RF)}$ denotes the multiplicative speckle noise,  which modifies the signal by multiplication rather than adding to it.}
In clinical practice, 2D B-mode \revision{US} images are derived by applying post-processing $\mathrm{P}_{log}(\cdot)$, including log compression and dynamic range adjustment, to the demodulated RF signals to improve visual clarity. Under this transformation, the speckle noise in B-mode images can be approximated as additive noise~\cite{coupe2009nonlocal}. Eq~(\ref{eq:speckle_noise}) can be rewritten as follows:
\begin{equation}~\label{eq:speckle_noise_log}
  \mathbf{Y} = \mathbf{X} + \mathbf{N}
\end{equation}
where $\mathbf{Y} =\mathrm{P}_{log}{(\mathbf{Y}_{(RF)})}$ denotes the observed B-mode image, $\mathbf{X} = \mathrm{P}_{log}{(\mathbf{X}_{(RF)})}$ is the underlying clean image, and $\mathbf{N} = \mathrm{P}_{log}{(\mathbf{N}_{(RF)})}$ corresponds to the speckle noise component in the B-mode domain.
Historically, the speckle noise model $p(\mathbf{Y}|\mathbf{X})$ has been approximated using various parametric distributions, such as the Nakagami, homodyned K, and Rayleigh families~\cite{christensen2024systematized}. However, due to the strong tissue-dependence and the fundamental difference between speckle and random noise, it remains challenging to define a universal probabilistic model that accurately captures the true speckle distribution $p(\mathbf{Y}|\mathbf{X})$ across different anatomical regions and imaging settings.

This implies that the statistical behaviour of speckle in B-mode images fundamentally differs from the independent noise assumptions commonly made in self-supervised denoising methods~\cite{lehtinen2018noise2noise, krull2019noise2void, huang2021neighbor2neighbor}. Unlike pixel-wise random noise (e.g., Gaussian, Poisson, or Salt-and-Pepper), speckle exhibits strong local spatial dependencies and large-scale granular patterns, making it substantially more difficult to model and suppress. Although it may appear visually similar to noise, speckle originates from the coherent interference of echoes scattered by sub-wavelength tissue structures, and is therefore not truly random. As noted in~\cite{dantas2005ultrasound, prince2006medical}, speckle does not qualify as ``noise" in the conventional sense used in natural image or CT/MRI processing. Nonetheless, for consistency and ease of understanding, we use the terms ``speckle” and ``noise”, as well as ``despeckling” and ``denoising”, interchangeably throughout this paper, while fully acknowledging their fundamental differences.

An intuitive comparison between US speckle and common pixel-wise noise types is shown in Fig.~\ref{fig:fig_nosie_comparision}. It can be seen that tissue-dependent speckle causes more severe degradation than other noise types. Moreover, degraded examples from the BSD68 dataset~\cite{MartinFTM01} illustrate that the detail loss induced by US speckle and related artefacts is nearly irreversible. This highlights the challenge of speckle reduction, as speckle is closely linked to the underlying tissue’s texture and structure. 
\subsection{Noise2Clean based on Clean Images}
\par
The fully supervised N2C paradigm has both noisy and clean images. With a large number of noisy-clean pairs $(\mathbf{Y},\mathbf{X})$, N2C tries to minimize the following loss in terms of $\theta$:
\begin{equation}~\label{eq:loss_N2C}
    \underset{\theta}{\arg \min }\ \mathbb{E}_{\mathbf{X}, \mathbf{Y}}\left\|f_\theta(\mathbf{Y})-\mathbf{X}\right\|_2^2
\end{equation}
where $f_\theta(\cdot)$ denotes the denoising network.
However, in many real-world cases, obtaining clean images is impractical, which significantly limits the applicability of this paradigm. 

\subsection{Noise2Noise based on Noisy Image Pairs}
\par
\revision{N2N}~\cite{lehtinen2018noise2noise} presents a pioneering approach to denoising without requiring clean reference images. 
Instead of training on noisy-clean pairs, N2N trains the network to predict one noisy observation $\mathbf{Y}$ from another independent noisy observation $\mathbf{Z}$ of the same underlying scene. Lehtinen et al. showed that under the zero-mean noise assumption, training a network to predict one noisy observation from another implicitly leads it to recover the underlying clean signal. The training objective is:

\begin{equation}~\label{eq:loss_N2N}
    \underset{\theta}{\arg \min }\ \mathbb{E}_{\mathbf{Y}, \mathbf{Z}}\left\|f_\theta(\mathbf{Y})-\mathbf{Z}\right\|_2^2
\end{equation}
where $f_\theta(\cdot)$ can adopt any CNN architecture, with U-Net~\cite{ronneberger2015u} commonly used due to its skip connections that help preserve fine structural details. 
With sufficient data, the denoising network $f_\theta(\cdot)$ can theoretically converge to the underlying clean image. 
However, this assumption does not hold for US speckle, which is neither zero-mean nor random. US speckle is inherently deterministic and highly tissue-dependent; repeated acquisitions of the same scene under identical settings produce nearly identical images. This violates the independence assumption required by N2N, making it inapplicable for speckle reduction in most commercial US systems.

\subsection{Nosie2Void based on Single Noisy Images}
\par
Compared to N2N~\cite{lehtinen2018noise2noise}, the training of N2V~\cite{krull2019noise2void} requires only single noisy images. In N2V~\cite{krull2019noise2void}, a CNN predicts the pixel value at the centre of a patch by using all surrounding pixels in the patch as inputs. By excluding the centre pixel of each patch, this design prevents the network from learning trivial identical mappings of the value at the centre of the input patch to the output, encouraging it to infer the underlying clean signal. This method is also intuitively termed to blind-spot network (BSN). The parameters of this network are optimised using the following equation:
\begin{equation}~\label{eq:loss_N2V_test}
    \underset{\theta}{\arg \min }\ \mathbb{E}_{\mathbf{Y}}\left\|f_\theta(\mathbf{Y}_p) - {Y_c} \right\|_2^2
\end{equation}
where $\mathbf{Y}_p$ represents a patch $\mathbf{Y}$ centered on pixel $Y_c$ by removing center pixel $Y_c$. For operational efficiency, N2V also introduces a masking scheme in which the center pixel’s value is replaced by a randomly selected value from the surrounding area to update $Y_c$. However, this method relies on two assumptions: (1) it assumes spatially independent noise within a single image, making it unsuitable for tissue-dependent noise; (2) it performs pixel-level denoising using a blind-spot network. 
These assumptions are in conflict with the characteristics of US speckle, as discussed in Sec.~\ref{sec:us_property}: its strong local spatial dependency violates the independent noise assumption, and each speckle affects a region far larger than a single pixel (see Fig.~\ref{fig:fig_nosie_comparision}), making BSN-based methods less ideal for speckle removal.

\section{Method}
\label{sec:methods}
\subsection{Overview of the Proposed Framework}
Although self-supervised denoising methods have rapidly progressed in recent years, they are mainly designed for random noise in natural images. As discussed in the previous section, the structured, tissue-dependent nature of US speckle makes these methods ineffective for US despeckling. So far, no self-supervised method can remove US speckle using only single noisy images without extra requirements. To tackle this challenge, we draw inspiration from low-rank approximation theory, which is widely used to separate structured signals from noise. Building on this foundation, we introduce our method that combines multi-scale perturbation with low-rank priors to achieve effective speckle reduction from single B-mode images.


\subsection{Low Rank Approximation}
\label{subsec:low_rank}
\revision{We first introduce a low-rank prior, leveraging the assumption that the underlying clean image $\mathbf{X}$ possesses a low-rank structure. This assumption has been widely applied in computer vision low-level tasks~\cite{wright2009robust,zheng2012practical,lu2016tensor, lefkimmiatis2023learning,sagheer2017ultrasound}.} One commonly used low-rank approximation method is Robust Principal Component Analysis (RPCA), which approximates the rank operation to separate the low-rank structure from sparse noise components, which is achieved by optimizing the following equation: 
\begin{equation}~\label{eq:lowRank}
    \min _{\mathbf{L}, \mathbf{S}}\mathrm{rank} (\mathbf{L})+\|\mathbf{S}\|_0, \quad \text { s.t. } \quad \mathbf{I}=\mathbf{L}+\mathbf{S}
\end{equation}
where $\mathrm{rank} (\mathbf{L})$ denotes the rank of the matrix to constraint low-rank property. $\|\mathbf{S}\|_0$ measures the sparsity of the matrix $\mathbf{S}$. 
In addition to RPCA, Singular Value Decomposition (SVD) is also commonly used to extract low-rank structures by retaining only the dominant singular values, effectively approximating the matrix with a lower-rank version while discarding high-frequency or noisy components.
The concept of low-rank approximation has also been widely applied in US imaging. In ~\cite{zhu2017non, sagheer2017ultrasound}, \revision{the clean tissue is treated as a low-rank signal}, while speckle is treated as a sparse component. Similarly, \cite{yang2024frequency} separates the low-rank clean US signal from intensity-focused US (HIFU) degraded images. More recently, low-rank approximation has been widely adopted in ultrafast US imaging for spatiotemporal clutter filter, where tissue signals are modelled as low-rank and blood flow as sparse components or high rank in spatiotemporal sequences \cite{demene2015spatiotemporal, chen2025ultrafast}.



\begin{figure}[ht!]
\centering
    \includegraphics[width=1\linewidth]{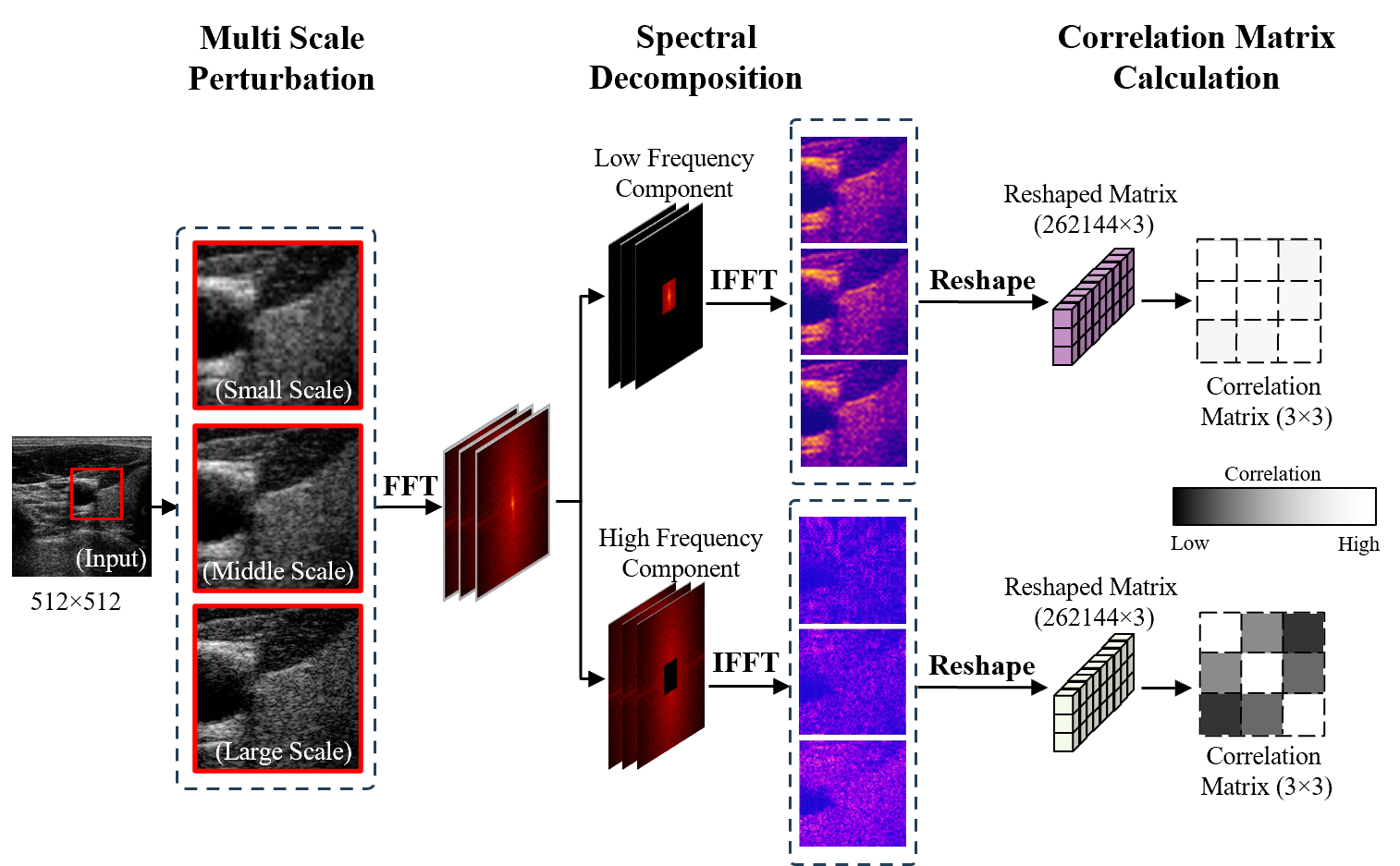}
    \caption{Frequency-domain analysis of \revision{image variants generated by MSP operation}. Low-frequency components show high cross-scale correlation, indicating shared low-rank tissue structure, while high-frequency components vary significantly due to scale-induced speckle differences.}
    \label{fig:fig_freq_analysis}
\end{figure}

\begin{figure*}[ht!]
  \centering
  \includegraphics[width=0.95\linewidth]{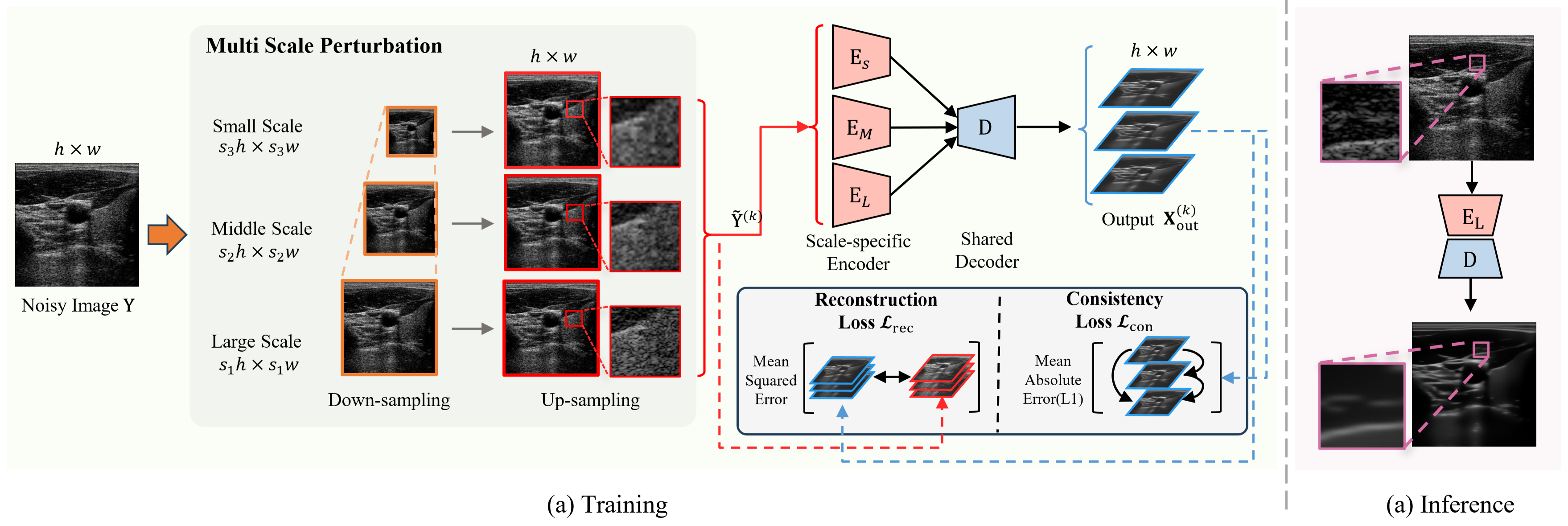}
  \caption{Overview of our proposed despeckling framework. (a) Overall training process. (b) Inference using the trained despeckling model. The network generates despeckled images directly from original US images.}
  \label{fig:network_overview_3}
\end{figure*}

\subsection{Implicit Low-Rank Approximation for US Speckle}
Classical low-rank approximation methods aim to explicitly decompose an image into low-rank and residual components, ideally isolating the clean structural signal from noise such as speckle. However, in US imaging, speckle noise is intricately entangled with anatomical structures, making such explicit decomposition methods, including RPCA or SVD, unreliable and difficult to apply. This motivates the need for an alternative approach that can implicitly separate the underlying low-rank content from speckle using only a single noisy observation, without access to clean references or temporal redundancy.


To tackle this, \revision{our approach also leverages the low-rank prior: the clean tissue $\mathbf{X}$ is assumed to lie on a low-rank subspace, while speckle noise $\mathbf{N}$ is more prominently expressed in the high-rank or sparse components.} To enable implicit separation of these components from just a single noisy observation $\mathbf{Y}$, we design a transformation called multi-scale perturbation (MSP), denoted as $\mathcal{P}_{k}(\cdot)$, where $k$ indicates the perturbation scale. MSP applies controlled downsampling–upsampling operations to generate multiple perturbed versions of the same noisy B-mode image $\mathbf{Y}$:
\begin{equation} \label{eq:updownSampling_2}
    \tilde{\mathbf{Y}}^{(k)} = \mathcal{P}_{k}(\mathbf{Y}) = \mathbf{X} + \mathbf{N}^{(k)}, \quad k = 1, 2, 3
\end{equation}
\revision{Specifically, this operation first downsamples the image by a factor of $s_k$ in both spatial dimensions (i.e., scaling the width and height by $s_k$), then upsamples it back to the original image size.}
The generated variants $\{\tilde{\mathbf{Y}}^{(k)}\}_{k=1}^3$ preserve the same anatomical structure $\mathbf{X}$, while exhibiting different speckle patterns $\mathbf{N}^{(k)}$ due to the scale-dependent distortions introduced by $\mathcal{P}_{k}(\cdot)$. As illustrated in Fig.~\ref{fig:network_overview_3}, \revision{after applying MSP, the boundaries of the thyroid remain visually consistent across different scales, while the speckle patterns in the homogeneous thyroid region vary significantly across the perturbed versions—effectively forming multiple images that share the same low-rank content but differ in speckle characteristics for the same scene.} In our implementation, we use three scale factors $ s_k \in \{1.0, 0.5, 0.25\} $, corresponding to $ k = 1, 2, 3 $, where each input is first downsampled by $ s_k $ and then upsampled back to the original resolution.

To investigate whether the MSP is able to introduce enough variation for the tissue speckle component, we perform a frequency-domain analysis on the three images after applying MSP with different scales. The low-rank tissue component is approximately extracted using the low-frequency bands, and the remaining high-frequency component is referred to as speckle. To this end, the Fourier transform (FFT) is applied to each variant (Fig.~\ref{fig:fig_freq_analysis}). \revision{Following ~\cite{yang2022source}, we isolate different frequency bands using the patch mask centered at the frequency image center}, and reconstruct corresponding sub-images via inverse FFT. In Fig.~\ref {fig:fig_freq_analysis}, the low-rank part shows high consistency while the high frequency exhibits a certain variation. This finding is also supported by the following correlation analysis across scales. This demonstrates the MSP can generate variants with common low-rank content but distinct non-structural components, including speckle. This lays the foundation for the following tissue-dependent speckle reduction.



\subsection{Network Architecture}

Building on the MSP-generated variants—each sharing a common low-rank anatomical structure but differing in speckle patterns—we design a disentanglement-inspired self-supervised learning framework. The goal is to extract the shared clean structure from these perturbed observations. Specifically, we adopt a multi-encoder architecture, where each scale-specific encoder $E_{k}$ processes its corresponding input $\tilde{\mathbf{Y}}^{(k)}$, and a shared decoder $D$ reconstructs the denoised output $\mathbf{X}^{(k)}_{\text{out}}$:
\begin{equation} \label{eq:encoder_decoder}
    \mathbf{X}^{(k)}_{\text{out}} = D \big( E_{k} ( \tilde{\mathbf{Y}}^{(k)} ) \big), \quad k = 1, 2, 3
\end{equation}
This architecture encourages the model to implicitly disentangle the consistent anatomical content from scale-sensitive speckle, as illustrated in Fig.~\ref{fig:network_overview_3}.

To optimize this probabilistic reconstruction model, we define a combined loss consisting of two complementary terms: a reconstruction loss and a consistency loss.
First, the reconstruction loss is defined as the mean squared error (MSE) between each reconstructed output and its corresponding input. The reconstruction loss encourages each output $\mathbf{X}^{(k)}_{\text{out}}$ to remain faithful to its corresponding perturbed input $\tilde{\mathbf{Y}}^{(k)}$, thereby preserving both structural and intensity fidelity at each scale:

\begin{equation}
\mathcal{L}_{\text{rec}}=\sum_{k=1}^{3}\left\|\mathbf{X}^{(k)}_{\text{out}}-\tilde{\mathbf{Y}}^{(k)}\right\|_2^2
\end{equation}

The consistency loss is designed to encourage the network to extract the stable low-rank structure that is invariant across multiple scale-perturbed views. While each perturbed input $\tilde{\mathbf{Y}}^{(k)}$ exhibits different speckle patterns, they all share the same underlying anatomical content. Therefore, their corresponding reconstructions $\mathbf{X}^{(k)}_{\text{out}}$ should be consistent with each other in terms of the shared tissue structure.
To enforce this, we compute the L1-norm between all pairs of reconstructed outputs. 
The L1-norm is chosen over MSE because it is more robust to outliers and sharp variations—properties often exhibited by speckle noise—making it more suitable for extracting consistent structural components. Unlike MSE, it does not force the network to overfit these variations, allowing the model to focus on stable, shared components across scales. This avoids the undesired averaging of scale-specific speckle variations, which would otherwise be smoothed out and retained in the final reconstruction.

The consistency loss is defined as:
\begin{equation}
\mathcal{L}_{\text {con}}=\sum_{k \neq l}\left\|\mathbf{X}^{(k)}_{\text{out}}-\mathbf{X}^{(l)}_{\text{out}}\right\|_1, \quad k,l \in \{1,2,3\}
\end{equation}

By jointly optimising these two losses, our self-supervised framework effectively extracts the underlying clean structure from multiple noisy variants in an implicit manner:
\begin{equation}~\label{eq:loss_sum}
  \mathcal{L} = \mathcal{L}_{rec} + \mathcal{L}_{con} 
\end{equation}

\section{Experiments}
\label{sec:experiment}

\subsection{Implementation Details}
In this study, both the encoder and decoder adopt structures similar to those in the Residual Autoencoder architecture~\cite{he2016deep}, although other common CNN designs could also be applied. Each encoder consists of four convolutional blocks followed by three residual blocks, while each decoder comprises four convolutional blocks and four residual blocks. A total of three encoders $\{E_{k}\}_{k=1,2,3}$ are used to process multi-scale inputs, all followed by a shared decoder $D$ for unified reconstruction. \revision{In total, the learnable parameters of the proposed model are about three million.} During training, we use the Adam optimizer with a learning rate of $0.001$, a batch size of $16$, and train for $3000$ epochs. The implementation is based on PyTorch 2.4 and was trained on an NVIDIA RTX 4080 Super GPU.

\subsection{Dataset Preparation}
\par
\subsubsection{Synthetic Datasets}
\par
We use the Matlab Ultrasound Toolbox (MUST)~\cite{perrot2021so, garcia2022simus, cigier2022simus} to simulate realistic ultrasound images from clean targets, employing a plane wave-based virtual transducer. To mitigate the poor quality often seen in single-angle plane wave imaging, we apply classic spatial compounding (SC), which fuses multiple images acquired at different tilt angles. This technique effectively suppresses speckle and enhances image clarity. Detailed simulation settings are listed in TABLE~\ref{tab:para_sim}.

\begin{table}[ht!]
    \centering
    \caption{
        Main simulation parameters, with the 7.50 MHz central frequency $f_{c}$ (marked in bold) used exclusively during training. The sampling frequency $f_{s}$ is consistently set to four times the central frequency $f_{c}$. Additionally, since plane wave imaging is used, there is no need to define a focusing depth $d_f$, which is required in commercial focused ultrasound imaging systems. Three additional central frequencies 3.75 MHz, 5.63 MHz, and 9.38 MHz are used for cross-configuration testing in Section~\ref{sec:cross_conf}.
    }
    \resizebox{0.37\textwidth}{!}{
    \begin{tabular}{lcc}
        \toprule
        \textbf{Parameters} & \textbf{Virtual Transducer} & \textbf{Unit}\\
        \midrule
          \multirow{1}{*}{Transducer Type} & linear & -\\
          \multirow{1}{*}{Transducer Model} & L12-3V & -\\
        \multirow{1}{*}{Central Frequency $f_{c}$}  &  3.75 / 5.63 / \textbf{7.50} /  9.38 & MHz \\
        \multirow{1}{*}{Sampling Frequency $f_{s}$} & $4 \times f_{c}$ & MHz\\
        \multirow{1}{*}{Focusing Depth $d_f$} & None & mm\\
        \multirow{1}{*}{Elements Number $n_e$} & 192 & -\\
        \multirow{1}{*}{Tilt Range $\theta_{\text{range}}$} & $-2^\circ$ to $2^\circ$ & $\text{degrees} (^\circ)$ \\
        \multirow{1}{*}{Plane Wave Number $n_c$} & 9 & -\\ 
        \bottomrule
        
    \end{tabular}
    }

    \label{tab:para_sim}
\end{table}

\par
The generation of synthetic noisy US images from a single tilt angle involves the following key steps:

\begin{enumerate} 
\item \textbf{Scatterer Creation}: Scatterers are initialized according to the gray-scale values of the clean target image, where intensity variations modulate scatterer intensity, mimicking heterogeneous tissue reflectivity.
\item \textbf{Acoustic Simulation}: Using the configured parameters of a virtual ultrasound transducer, the acoustic pressure field is simulated to model the interaction between ultrasound waves and scatterers.
\item \textbf{RF Signal Processing}: The resulting raw  RF signals are processed through beamforming, demodulation, and time-gain compensation (TGC) to prepare for image formation.
\item \textbf{B-mode Image Formation}: The processed signals are log-compressed to generate the final B-mode ultrasound image, producing a grayscale representation consistent with real clinical outputs. 
\end{enumerate}

\revision{
Each synthetic dataset contains paired clean and speckled B-mode images at a resolution of 512×512 pixels. To enhance the diversity of the data type, three types of synthetic data are computed, where S-I includes heavy artifacts beyond speckle, generated using a larger tilt range $\theta_{\text{range}}$; S-II and S-III have fewer artifacts and are generated using the parameters listed in TABLE~\ref{tab:para_sim}, with S-II sharing similar shapes with S-I, and S-III featuring different shapes. Each image was generated in 20–30 minutes. The total images for each group has been summarized in Table~\ref{tab:data_overview}. 
}

\begin{table}[h!]
    \scriptsize
    \centering
    \caption{\revision{Visual illustration of the synthetic datasets in three different simulated settings and carotid US images from Clarius. Synthetic datasets have simulated clean and noisy images. The carotid US images obtained from volunteers only have speckled images without clean data.}}
    \resizebox{0.46\textwidth}{!}{
    \begin{tabular}{>{\centering\arraybackslash}m{1.5cm} 
                    >{\centering\arraybackslash}m{1.2cm} 
                    >{\centering\arraybackslash}m{1.2cm} 
                    >{\centering\arraybackslash}m{1.2cm}
                    >{\centering\arraybackslash}m{1.2cm}}
        \toprule
        \textbf{Dataset} & \textbf{S-I Data} & \textbf{S-II Data} & \textbf{S-III Data} & \textbf{Carotid Images}\\
        \midrule
        \textbf{Train} & 154 & 128 & 148 & 419\\
        \midrule
        \textbf{Test} & 39 & 32 & 37 & 104\\
        \midrule
        \textbf{Clean Target} & 
            \includegraphics[width=0.08\textwidth]{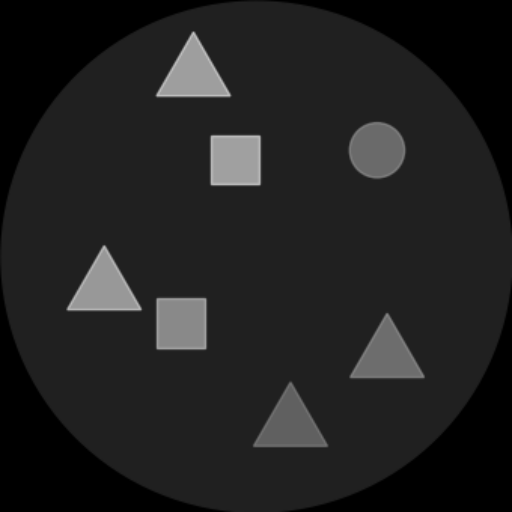} & 
            \includegraphics[width=0.08\textwidth]{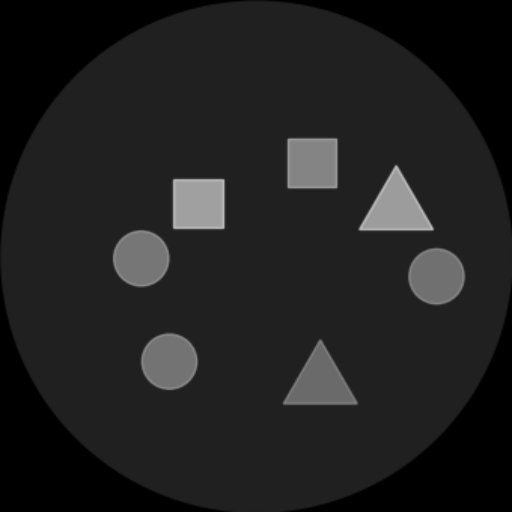} & 
            \includegraphics[width=0.08\textwidth]{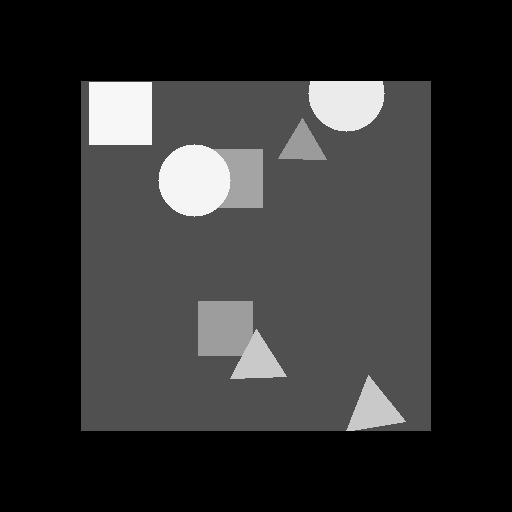} &
            \includegraphics[width=0.08\textwidth]{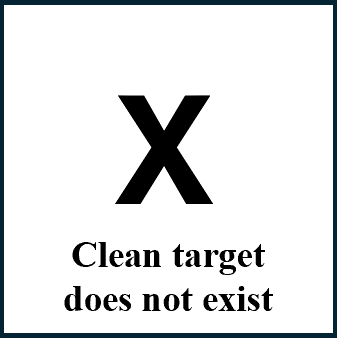}\\
        \midrule
        \textbf{Noisy Image} & 
            \includegraphics[width=0.08\textwidth]{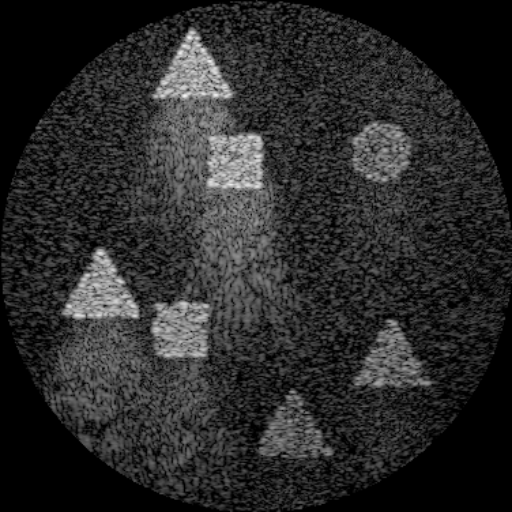} & 
            \includegraphics[width=0.08\textwidth]{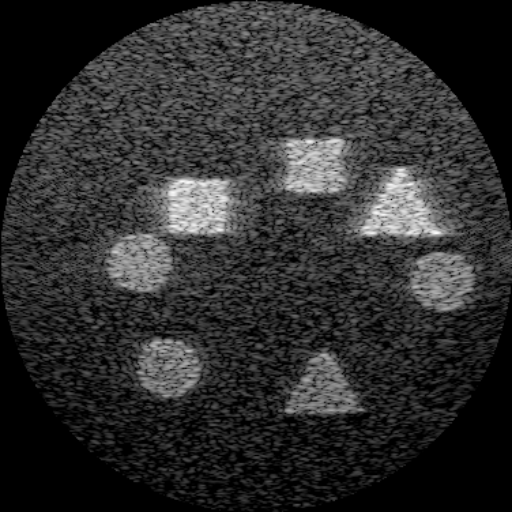} & 
            \includegraphics[width=0.08\textwidth]{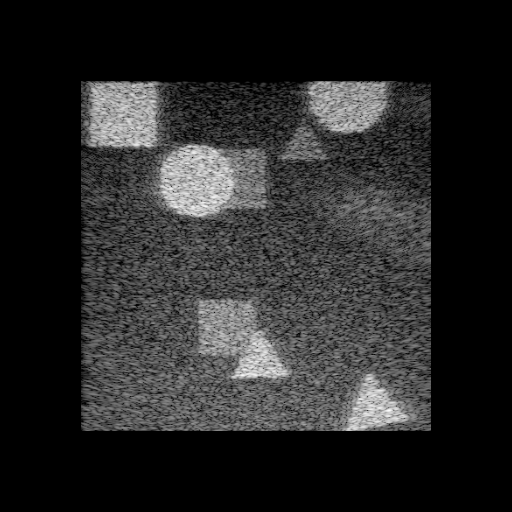} &
            \includegraphics[width=0.08\textwidth]{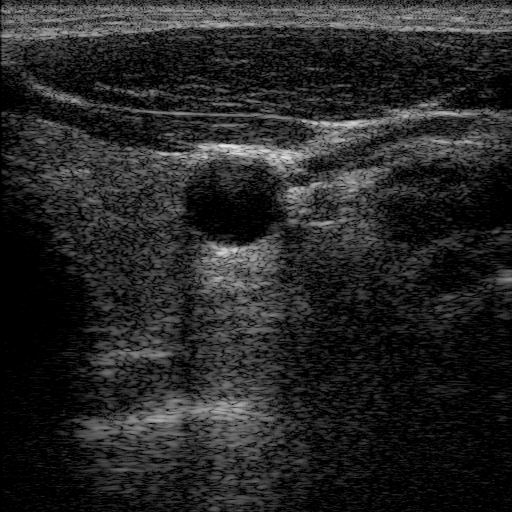} \\

        \bottomrule
    \end{tabular}
    }

    \label{tab:data_overview}
\end{table}

\subsubsection{In-vivo Datasets}
\par
To train and evaluate the proposed method on clinical US data, we collected carotid B-mode images using a Clarius L7 portable ultrasound scanner. The noisy images were acquired directly via the device’s API for envelope collection. Detailed acquisition settings are summarised in TABLE~\ref{tab:para_device}. The training set consists of 419 2D US images at a resolution of 512×512 pixels, collected from five healthy volunteers. An independent test set of 104 images was acquired from two additional volunteers to assess generalisation performance.

\begin{table}[ht!]
    \centering
    \caption{Ultrasound settings used across three different devices for real-case experiments.
    Clarius (marked in bold) serves as the primary device for both Despeckling Performance and Ablation Study evaluations. Data from Siemens and Cephasonics are used exclusively for Cross-Device Validation in Section~\ref{sec:cross_device}.}
    \resizebox{0.42\textwidth}{!}{
    \begin{tabular}{llllc}
        \toprule
        \textbf{Parameters} & \textbf{Clarius} & \textbf{Siemens} & \textbf{Cephasonics} & \textbf{Unit} \\
        \midrule
        \multirow{1}{*}{Transducer Type} 
            & \textbf{linear} & linear & linear & -\\
        \multirow{1}{*}{Transducer Model} 
            & \textbf{HD L7} & 12L3 & CPLA12875 & -\\
        \multirow{1}{*}{Central Frequency $f_{c}$} 
            & \textbf{10.00} & 6.20 & 7.60 & MHz\\
        \multirow{1}{*}{Sampling Frequency $f_{s}$} 
            & \textbf{-} & - &  40.00 & MHz\\
        \multirow{1}{*}{Focusing Depth $d_f$} 
            & \textbf{15.00} & 20.00 & 20.00 & mm\\
        \multirow{1}{*}{Imaging Depth $d_i$} 
            & \textbf{40.00} & 40.00 & 40.00 & mm\\
        \multirow{1}{*}{Elements Number $n_e$}  
            & \textbf{192} & 192 & 128 & -\\
        \bottomrule
    \end{tabular}
    }

    \label{tab:para_device}
\end{table}

\begin{table*}[ht!]
    \centering
    \caption{Quantitative comparison of different methods for US speckle reduction in terms of PSNR, SSIM, Homogeneity(Hom), and \revision{LPIPS}. The best-performing method for each metric, trained without clean images, is highlighted in bold. The supervised method results are marked in a gray shade. Note that for clinical carotid images, the N2N and supervised Noise2True methods are not applicable.}
    \resizebox{1\textwidth}{!}{
    \begin{tabular}{lcccccc}
        \toprule
        \multirow{2}{*}{\textbf{Type of supervision}} & \multirow{2}{*}{\textbf{Method}} & \textbf{S-I Data} & \textbf{S-II Data} & \textbf{S-III Data} & \textbf{Carotid} & \multirow{2}{*}{\textbf{Time}}\\
        & & PSNR($\uparrow$)/SSIM($\uparrow$)/Hom($\uparrow$)/\revision{LPIPS($\downarrow$)} & PSNR($\uparrow$)/SSIM($\uparrow$)/Hom($\uparrow$)/\revision{LPIPS($\downarrow$)} & PSNR($\uparrow$)/SSIM($\uparrow$)/Hom($\uparrow$)/\revision{LPIPS($\downarrow$)} & Hom($\uparrow$) & \\
        
        \midrule
        \multirow{4}{*}{Non-learning based} 
            & SRAD~\cite{yu2002speckle} & \textbf{19.40} / 0.734 / 0.763 / \revision{0.411} & \textbf{19.57} / 0.735 / 0.768 / \revision{0.419} & 20.81 / 0.880 / 0.874 / \revision{0.246} & 0.705 &  6 s.\\
            & NLM~\cite{buades2005non} & \textbf{19.40} / 0.736 / 0.714 / \revision{0.417} & 19.52 / 0.738 / 0.740 / \revision{0.428} & 20.94 / 0.881 / 0.839 / \revision{0.310} & 0.691 &  80 ms.\\
            & BM3D~\cite{dabov2007image} & 19.26 / 0.684 / 0.691 / \revision{0.397} & 19.34 / 0.679 / 0.694 / \revision{0.407} & 21.01 / 0.875 / 0.844 / \revision{0.245} & 0.656 & 3 s.\\
            & OBNLM~\cite{coupe2009nonlocal} & 18.84 / 0.703 / 0.577 / \revision{0.470} & 18.88 / 0.702 / 0.621 / \revision{0.484} & 21.49 / 0.885 / 0.774 / \revision{0.299} & 0.460 &  12 s.\\
        \midrule

        Supervised & \multirow{2}{*}{Noise2True} & \multirow{2}{*}{\colorbox{gray!20}{23.33 / 0.936 / 0.910 / \revision{0.059}}} &  \multirow{2}{*}{\colorbox{gray!20}{23.25 / 0.937 / 0.923 / \revision{0.049}}} &  \multirow{2}{*}{\colorbox{gray!20}{27.66 / 0.973 / 0.941 / \revision{0.062}}} & \multirow{2}{*}{$\times$} &  \multirow{2}{*}{6 ms.}\\
        (Noisy-Clean Pairs) & & & & & & \\
        \midrule
 
        Self-supervised & \multirow{2}{*}{N2N~\cite{lehtinen2018noise2noise}} & \multirow{2}{*}{18.05 / 0.705 / 0.683 / \revision{0.369}} & \multirow{2}{*}{17.74 / 0.691 / 0.689 / \revision{0.410}} & \multirow{2}{*}{21.99 / 0.895 / 0.820 / \revision{0.240}} & \multirow{2}{*}{$\times$} & \multirow{2}{*}{6 ms.} \\
        (Noisy-Noisy Pairs) & & & & & & \\
        \midrule

        \multirow{7}{*}{\begin{tabular}{@{}l@{}}Self-supervised\\(Single Noisy Images)\end{tabular}} 
        & DIP(*)~\cite{ulyanov2018deep} & 18.87 / 0.543 / 0.515 / \revision{0.484} & 18.93 / 0.523 / 0.531 / \revision{0.501} & 20.98 / 0.831 / 0.747 / \revision{0.323} & 0.436 & $\times$\\
        & N2V~\cite{krull2019noise2void} & 17.62 / 0.311 / 0.312 / \revision{0.561} & 17.55 / 0.314 / 0.339 / \revision{0.551} & 19.87 / 0.613 / 0.573 / \revision{0.388} & 0.215 & 6 ms.\\
        & N2S(*)~\cite{batson2019noise2self} &  17.32 / 0.303 / 0.308 / \revision{0.549} & 17.64 / 0.310 / 0.339 / \revision{0.545} & 19.69 / 0.587 / 0.576 / \revision{0.399} & 0.227 & $\times$\\
        & N2S~\cite{batson2019noise2self} & 18.14 / 0.346 / 0.316 / \revision{0.543} & 17.64 / 0.344 / 0.330 / \revision{0.547} & 20.41 / 0.657 / 0.573 / \revision{0.399} & 0.184 & 6 ms.\\
        & Neighbor2Neighbor~\cite{huang2021neighbor2neighbor} & 17.63 / 0.319 / 0.310 / \revision{0.556} & 17.79 / 0.321 / 0.339 / \revision{0.549} & 20.19 / 0.637 / 0.574 / \revision{0.384} & 0.213 & 6 ms.\\
        & ZS-N2N(*)~\cite{mansour2023zero} & 17.58 / 0.315 / 0.312 / \revision{0.557} & 17.71 / 0.318 / 0.340 / \revision{0.550} & 20.04 / 0.625 / 0.576 / \revision{0.384} & 0.214 & $\times$\\
        & \revision{Speckle2Self (our)} & 18.59 / \textbf{0.771} / \textbf{0.803} / \revision{\textbf{0.231}} & 18.94 / \textbf{0.777} / \textbf{0.792} / \revision{\textbf{0.233}} & \textbf{22.90} / \textbf{0.922} / \textbf{0.876} / \revision{\textbf{0.173}} & \textbf{0.725} & 7 ms. \\

        \bottomrule
    \end{tabular}
    }

    \label{tab:quant_benchmark_results}
\end{table*}

\begin{figure}[htbp!]
\centering
    \includegraphics[width=0.98\linewidth]{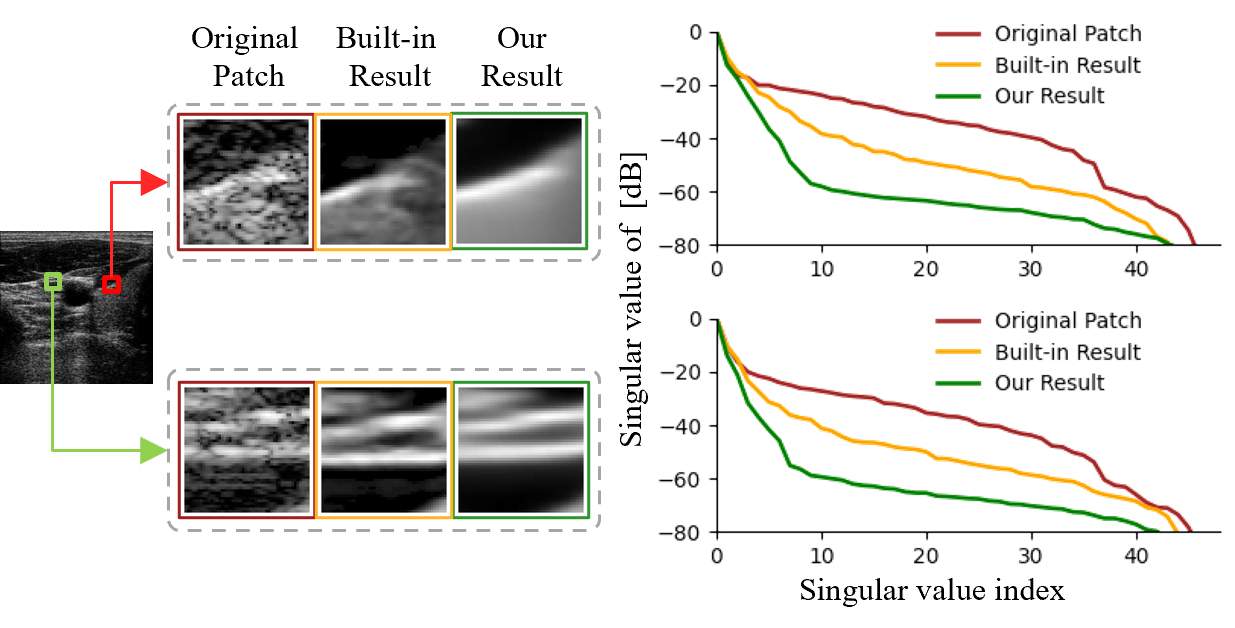}
    \caption{SVD-based singular value spectra of two image patches (top: homogeneous/edge, bottom: inhomogeneous/edge). The denoised patches show faster singular value decay, confirming enhanced low-rank characteristics.}
    \label{fig:low_rank_prof}
\end{figure}

\subsection{Low-rank Enhancement Performance} 

To intuitively demonstrate the ability of the proposed method to consistently preserve low-rank information, we conduct a Singular Value Decay analysis on three patches extracted from identical locations in three images: the original noisy image, the image denoised with built-in filters, and the image denoised using our proposed \revision{Speckle2Self}. Each patch has a size of $48\times48$ pixels. Two patches are selected — one from a relatively homogeneous region (thyroid) and another from a structurally complex area, as indicated by the green and red boxes in Fig.\ref{fig:low_rank_prof}. \revision{SVD} is performed on each patch before and after despeckling, with the singular values sorted in descending order. As shown in the curve plots in Fig.\ref{fig:low_rank_prof}, the Speckle2Self results exhibit a significantly faster singular value decay compared to their noisy and Clarius built-in counterparts. This indicates a stronger low-rank structure after processing, demonstrating the model’s ability to effectively suppress incoherent speckle while preserving critical tissue details. Notably, even in the structurally complex region, dominant singular components are retained, confirming that key anatomical structures are well preserved during speckle reduction.


\subsection{Comparisons with State-of-the-Arts} \label{sec:SOTA}
\subsubsection{Compared Methods}
\par
To evaluate the proposed method, we compared \revision{Speckle2Self} with classical denoising algorithms, including BM3D~\cite{dabov2007image}, NLM~\cite{buades2005non}, OBNLM~\cite{coupe2009nonlocal}, and SRAD~\cite{yu2002speckle}. Additionally, we tested popular self-supervised methods trained on datasets, such as N2N~\cite{lehtinen2018noise2noise}, N2S~\cite{batson2019noise2self}, N2V~\cite{krull2019noise2void}, and Neighbor2Neighbor~\cite{huang2021neighbor2neighbor}. We also included zero-shot self-supervised methods trained on single image, including DIP(*)~\cite{ulyanov2018deep}, N2S(*)~\cite{krull2019noise2void}, and ZS-N2N(*)~\cite{mansour2023zero}, as well as the full-supervised method Noise2True. The star * denotes that the marked method will need to be retrained in the zero shot self-supervised mode for each image. For traditional methods, we need to carefully tune the parameters to have optimized performance on our dataset. For BM3D~\cite{dabov2007image}, we directly adopt its official implementation, NLM~\cite{buades2005non} from OpenCV implementation, OBNLM from~\href{https://github.com/Xingorno/Optimized-Bayesian-Nonlocal-means-with-block-OBNLM}{OBNLM} and SRAD from~\href{https://github.com/Xingorno/Speckle-Reducing-Anisotropic-Diffusion-SRAD}{SRAD}.



\subsubsection{Evaluation Metrics}
\par
To assess despeckling performance, we used Peak Signal-to-Noise Ratio (PSNR) and Structural Similarity Index Measure (SSIM), two widely adopted metrics in image quality assessment and various inverse problems. \revision{Additionally, the Learned Perceptual Image Patch Similarity (LPIPS)~\cite{zhang2018unreasonable} metric was included to provide a more comprehensive evaluation of perceptual quality.} These full-reference metrics were applied to synthetic ultrasound datasets where clean ground truth images are available. However, for real carotid US images, where clean targets are inherently unavailable, full-reference metrics like PSNR and SSIM cannot be employed. In such cases—particularly in homogeneous regions like the thyroid, where speckle noise manifests as grainy texture and severely degrades image quality—it becomes necessary to rely on no-reference evaluation. To this end, we utilised the homogeneity measure from the Grey Level Co-occurrence Matrix (GLCM)~\cite{glcm}, a widely used texture analysis technique that quantifies local smoothness and structural regularity without the need for clean reference images.

\begin{figure*}[ht!]
    \small
    \centering
    \setlength{\tabcolsep}{1.5pt} 
    \renewcommand{\arraystretch}{1} 
    \resizebox{1\textwidth}{!}{
    \begin{tabular}{ccccccc}
        \textbf{Ground Truth} & 
        \textbf{Noisy Input} & 
        \textbf{BM3D} & 
        \textbf{NLM} & 
        \textbf{OBNLM} & 
        \textbf{SRAD} &
        \textbf{Noise2True} \\

        \begin{overpic}[width=0.14\textwidth]{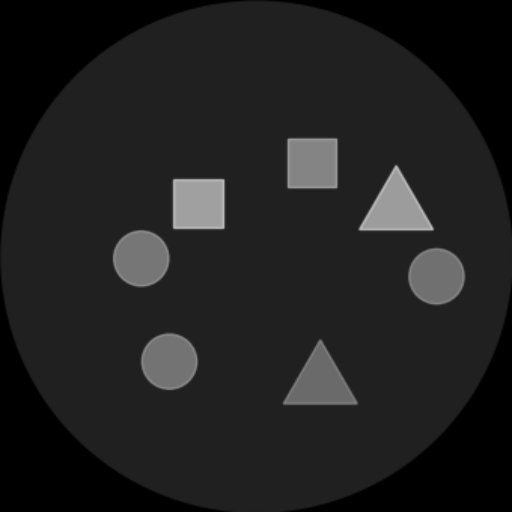} 
        \end{overpic}&
        
        \begin{overpic}[width=0.14\textwidth]{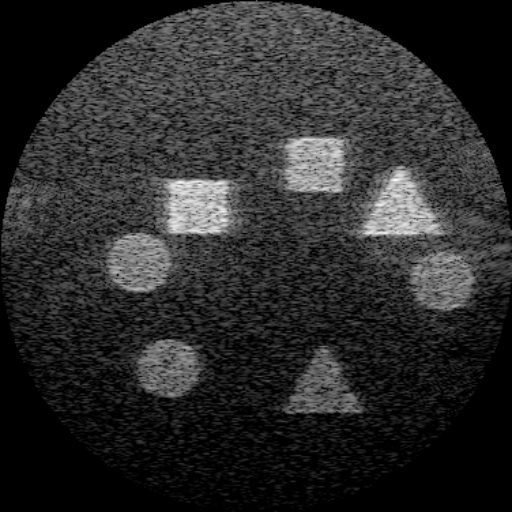} 
        \put(3, 5){\textbf{\textcolor{white}{\scriptsize SSIM: 0.319}}}
        \end{overpic}&
        
        \begin{overpic}[width=0.14\textwidth]{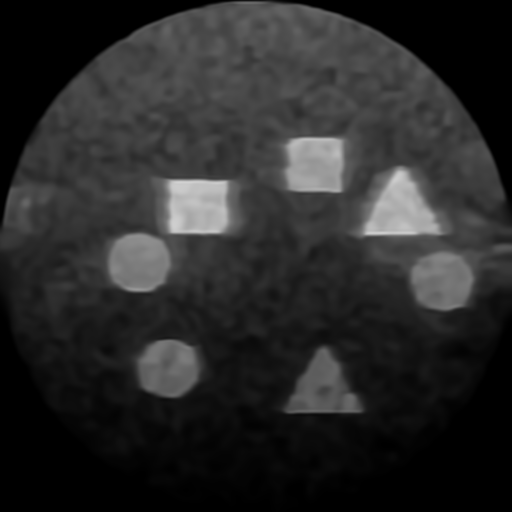} 
        \put(3, 5){\textbf{\textcolor{white}{\scriptsize SSIM: 0.679}}}
        \end{overpic}&
        
        \begin{overpic}[width=0.14\textwidth]{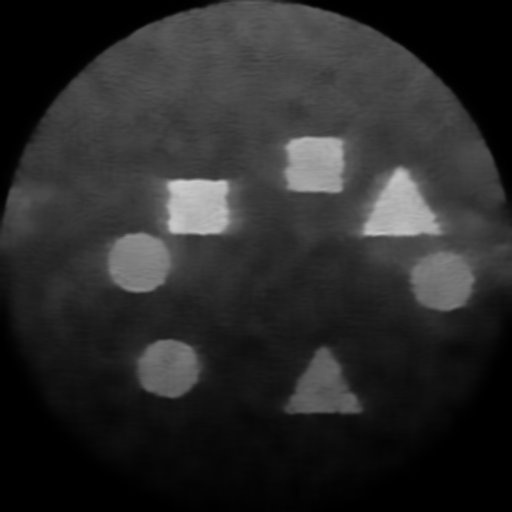} 
        \put(3, 5){\textbf{\textcolor{white}{\scriptsize SSIM: 0.738}}}
        \end{overpic}&
        
        \begin{overpic}[width=0.14\textwidth]{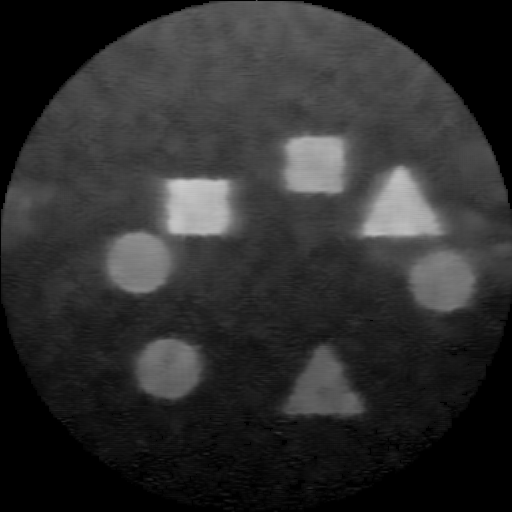} 
        \put(3, 5){\textbf{\textcolor{white}{\scriptsize SSIM: 0.702}}}
        \end{overpic}&
        
        \begin{overpic}[width=0.14\textwidth]{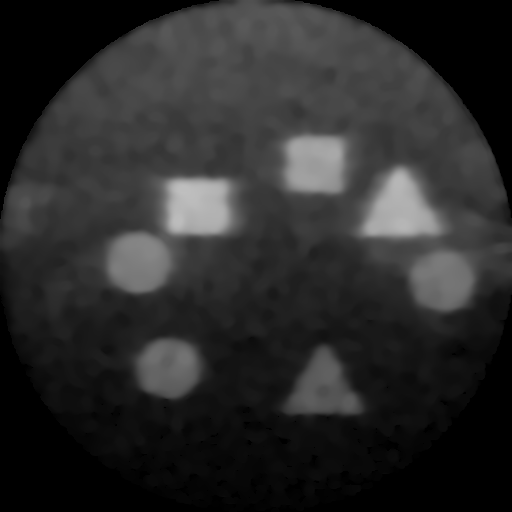} 
        \put(3, 5){\textbf{\textcolor{white}{\scriptsize SSIM: 0.735}}}
        \end{overpic}&
        
        \begin{overpic}[width=0.14\textwidth]{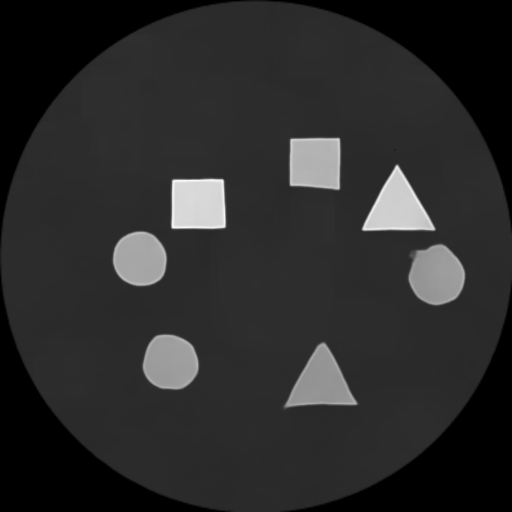} 
        \put(3, 5){\textbf{\textcolor{white}{\scriptsize SSIM: 0.937}}}
        \end{overpic}\\

        \textbf{Our Method} &  
        \textbf{N2N} & 
        \textbf{N2V} & 
        \textbf{N2S} & 
        \textbf{DIP(*)} &
        \textbf{ZS-N2N(*)} &
        \textbf{Neigh2Neigh}\\

        \begin{overpic}[width=0.14\textwidth]{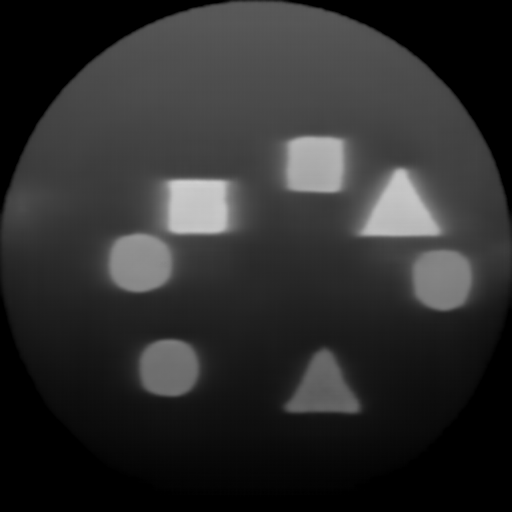} 
        \put(3, 5){\textbf{\textcolor{white}{\scriptsize SSIM: 0.777}}}
        \end{overpic}&
        \begin{overpic}[width=0.14\textwidth]{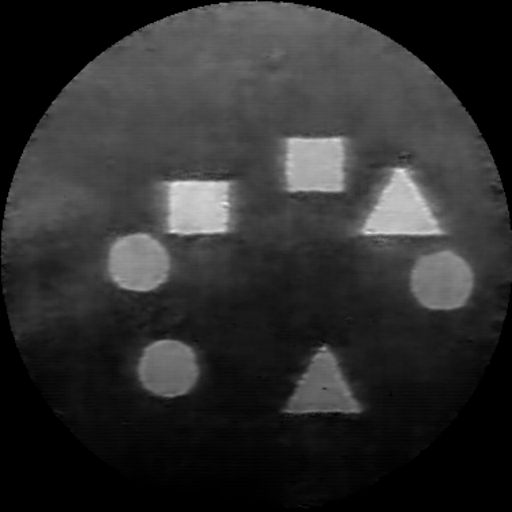} 
        \put(3, 5){\textbf{\textcolor{white}{\scriptsize SSIM: 0.691}}}
        \end{overpic}&
        \begin{overpic}[width=0.14\textwidth]{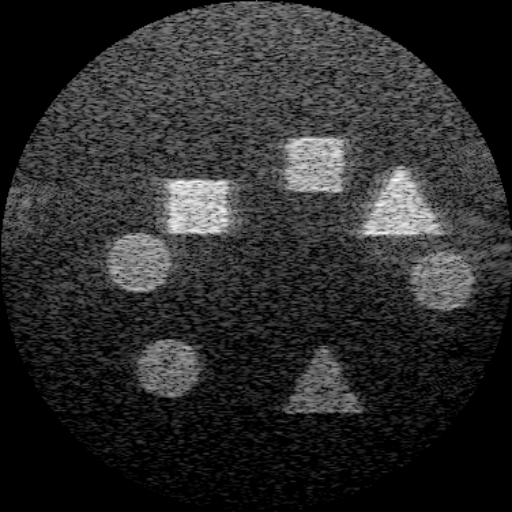} 
        \put(3, 5){\textbf{\textcolor{white}{\scriptsize SSIM: 0.314}}}
        \end{overpic}&
        \begin{overpic}[width=0.14\textwidth]{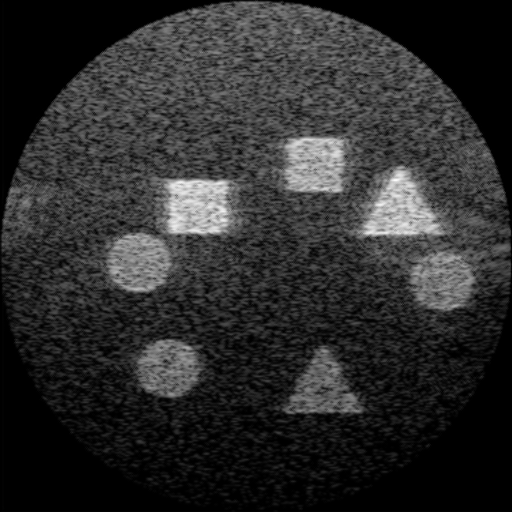} 
        \put(3, 5){\textbf{\textcolor{white}{\scriptsize SSIM: 0.344}}}
        \end{overpic}&
        \begin{overpic}[width=0.14\textwidth]{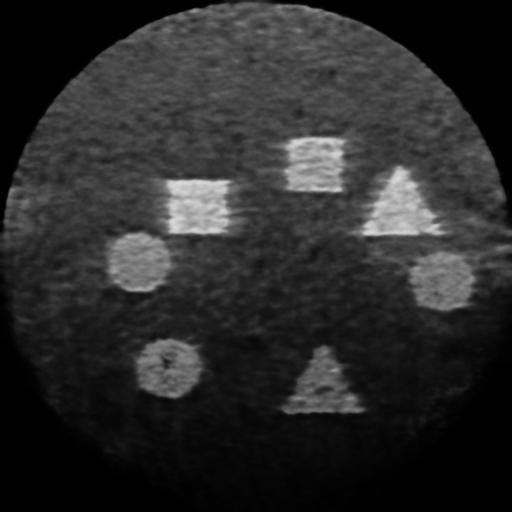} 
        \put(3, 5){\textbf{\textcolor{white}{\scriptsize SSIM: 0.523}}}
        \end{overpic}&
        \begin{overpic}[width=0.14\textwidth]{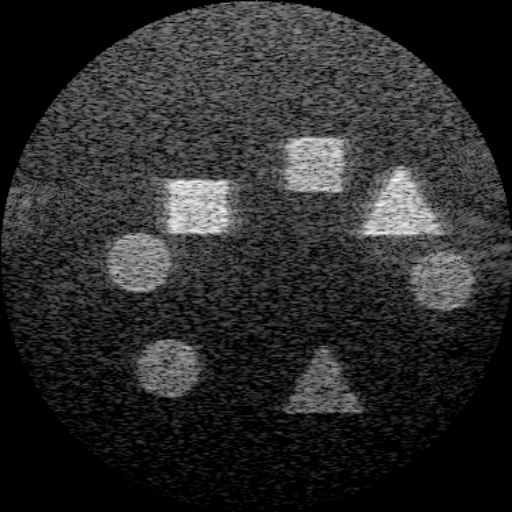} 
        \put(3, 5){\textbf{\textcolor{white}{\scriptsize SSIM: 0.318}}}
        \end{overpic}&
        \begin{overpic}[width=0.14\textwidth]{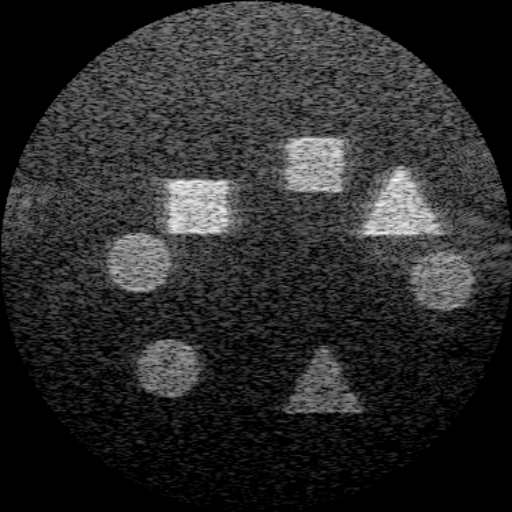} 
        \put(3, 5){\textbf{\textcolor{white}{\scriptsize SSIM: 0.321}}}
        \end{overpic}\\

    \end{tabular}
    }
    \caption{Despeckling results and average SSIM values on Synthetic S-II dataset.}
    \label{img:comparison_synthetic}
\end{figure*}

\begin{figure*}[ht!]
    \small
    \centering
    \setlength{\tabcolsep}{1.5pt} 
    \renewcommand{\arraystretch}{1} 
    \resizebox{1\textwidth}{!}{
    \begin{tabular}{ccccccc}
        \textbf{Ground Truth} & 
        \textbf{Noisy Input} & 
        \textbf{BM3D} & 
        \textbf{NLM} & 
        \textbf{OBNLM} & 
        \textbf{SRAD} &
        \textbf{Noise2True} \\
        
        \begin{overpic}[width=0.14\textwidth]{images/compare_imgs_CCA/clean_sign.png} 
        \end{overpic}&
        \begin{overpic}[width=0.14\textwidth]{images/compare_imgs_CCA/input.png} 
        \put(3, 5){\textbf{\textcolor{white}{\scriptsize Homogeneity: 0.237}}}
        \end{overpic}&
        \begin{overpic}[width=0.14\textwidth]{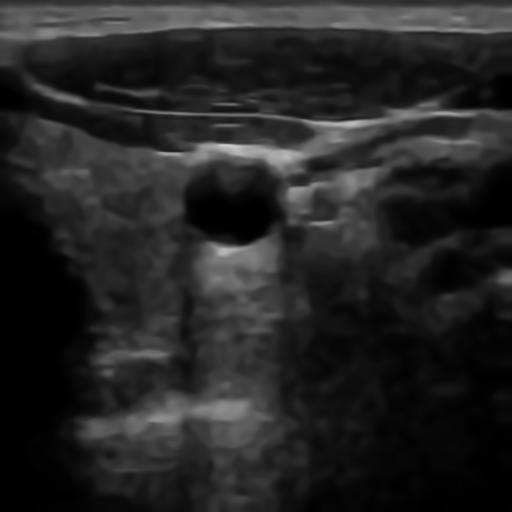} 
        \put(3, 5){\textbf{\textcolor{white}{\scriptsize Homogeneity: 0.656}}}
        \end{overpic}&
        \begin{overpic}[width=0.14\textwidth]{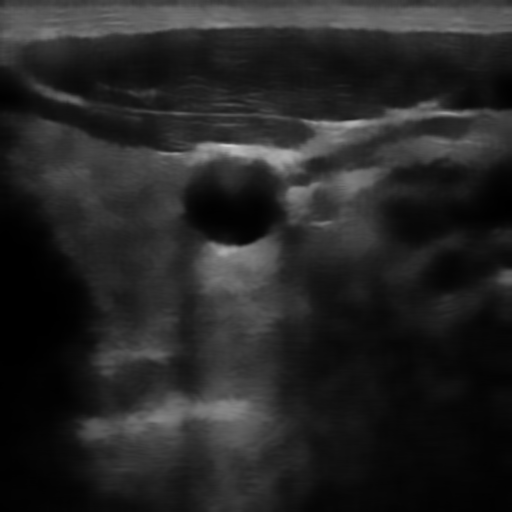} 
        \put(3, 5){\textbf{\textcolor{white}{\scriptsize Homogeneity: 0.691}}}
        \end{overpic}&
        \begin{overpic}[width=0.14\textwidth]{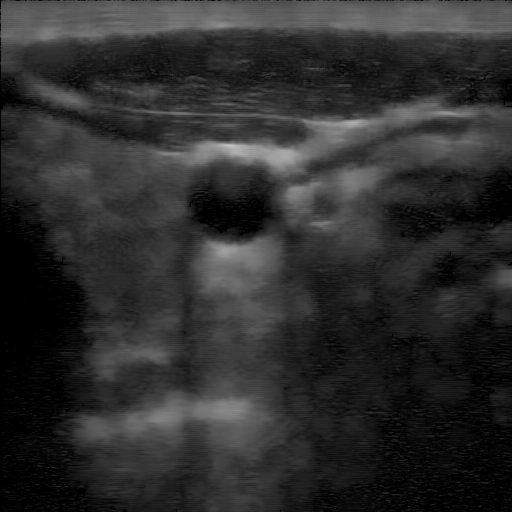} 
        \put(3, 5){\textbf{\textcolor{white}{\scriptsize Homogeneity: 0.460}}}
        \end{overpic}&
        \begin{overpic}[width=0.14\textwidth]{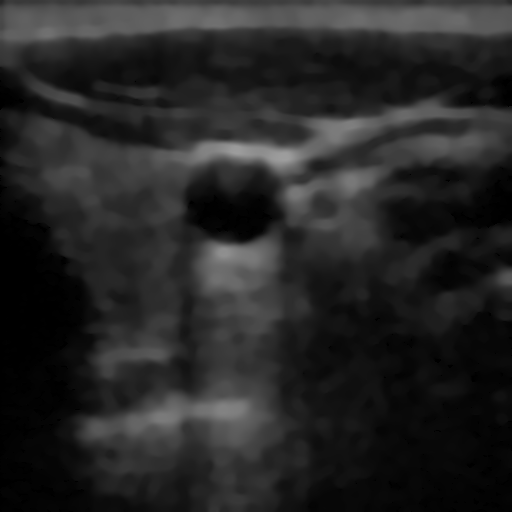} 
        \put(3, 5){\textbf{\textcolor{white}{\scriptsize Homogeneity: 0.705}}}
        \end{overpic}&
        \begin{overpic}[width=0.14\textwidth]{images/compare_imgs_CCA/clean_sign.png} 
        \end{overpic}\\

        \textbf{Our Method} & 
        \textbf{N2N} & 
        \textbf{N2V} & 
        \textbf{N2S} & 
        \textbf{DIP(*)} &
        \textbf{ZS-N2N(*)} &
        \textbf{Neigh2Neigh}\\ 

        \begin{overpic}[width=0.14\textwidth]{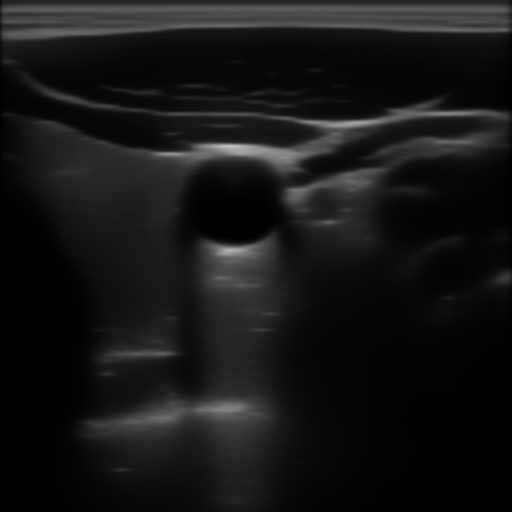} 
        \put(3, 5){\textbf{\textcolor{white}{\scriptsize Homogeneity: 0.725}}}
        \end{overpic}&
        \begin{overpic}[width=0.14\textwidth]{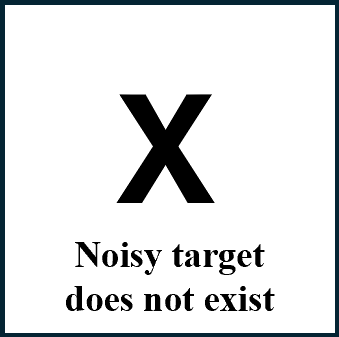} 
        \end{overpic}&
        \begin{overpic}[width=0.14\textwidth]{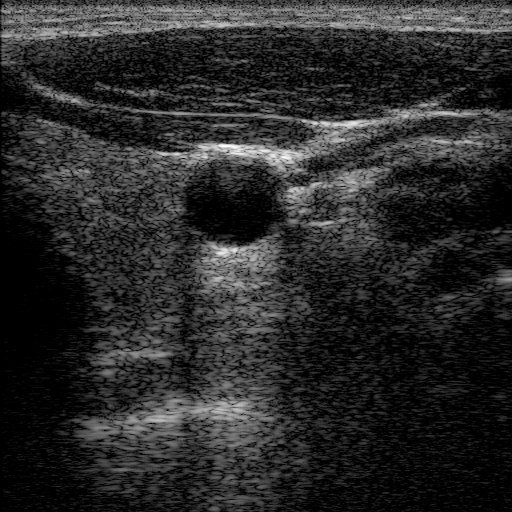} 
        \put(3, 5){\textbf{\textcolor{white}{\scriptsize Homogeneity: 0.215}}}
        \end{overpic}&
        \begin{overpic}[width=0.14\textwidth]{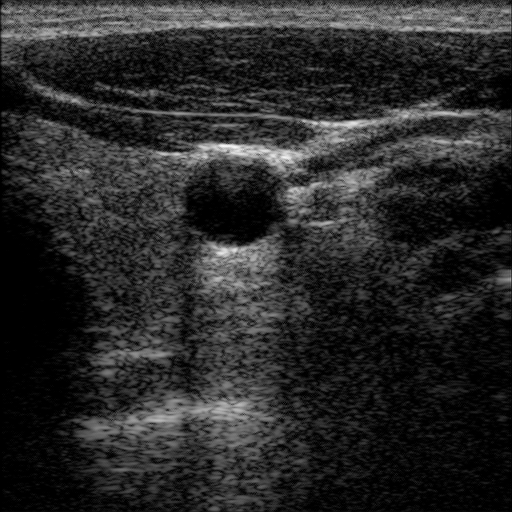} 
        \put(3, 5){\textbf{\textcolor{white}{\scriptsize Homogeneity: 0.227}}}
        \end{overpic}&
        \begin{overpic}[width=0.14\textwidth]{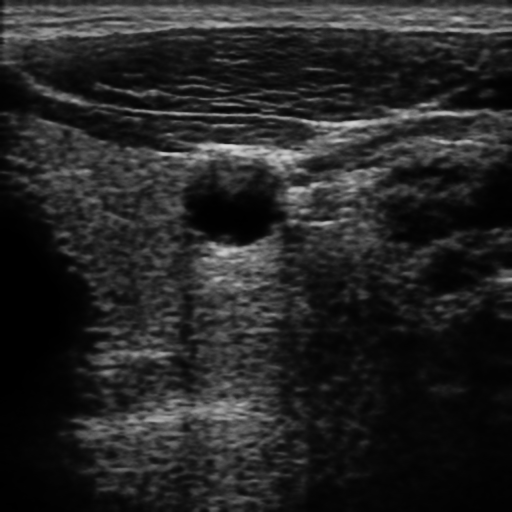} 
        \put(3, 5){\textbf{\textcolor{white}{\scriptsize Homogeneity: 0.436}}}
        \end{overpic}&
        \begin{overpic}[width=0.14\textwidth]{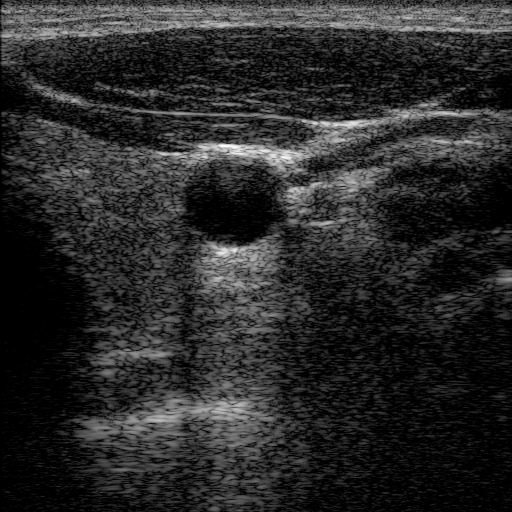} 
        \put(3, 5){\textbf{\textcolor{white}{\scriptsize Homogeneity: 0.214}}}
        \end{overpic}&
        \begin{overpic}[width=0.14\textwidth]{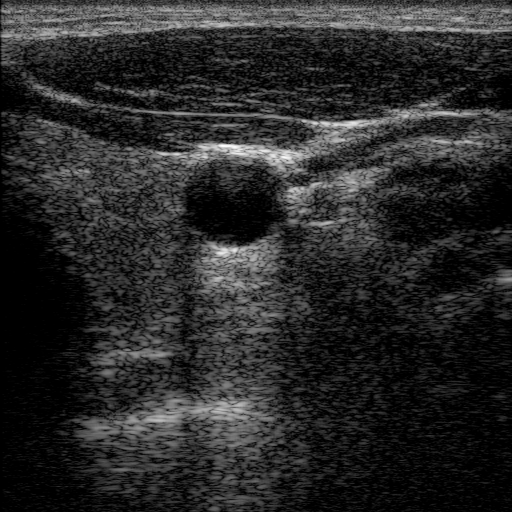} 
        \put(3, 5){\textbf{\textcolor{white}{\scriptsize Homogeneity: 0.213}}}
        \end{overpic}\\

    \end{tabular}
    }
    \caption{Despeckling results and average homogeneity values on carotid US dataset acquired from volunteers.}
    \label{img:comparison_CCA}
\end{figure*}

\subsubsection{Despeckling Performance on Synthetic Data}
The quantitative results computed using different methods across three synthetic datasets are summarized in TABLE~\ref{tab:quant_benchmark_results}. The supervised Noise2True method achieves the highest performance across all three metrics on all synthetic datasets, underscoring the advantages of supervised training when clean target images are available. However, as emphasized earlier, clean images are generally not obtainable in real scenarios. 

Without a clean reference image, our method significantly outperforms conventional US speckle denoising techniques and learning-based methods across all metrics, except for PSNR on the S-I/S-II dataset. The best results in SSIM, \revision{LPIPS,} and homogeneity show that the proposed method achieves strong structural preservation while effectively reducing speckle noise. Notably, non-learning-based methods also demonstrate significant effectiveness in US speckle denoising in comparison with learning-based methods (except ours). This is because, prior to our method, the self-supervised learning-based approaches mainly relied on the assumption of a pixel-level noise prior and independent noise distribution, which does not hold for tissue-dependent speckle in US imaging. In contrast, traditional methods often make weaker and less restrictive assumptions about noise distribution, allowing them to remain partially effective even in the presence of speckle.

\par
\revision{
Since the PSNR is often used to highlight the pixel-by-pixel similarity, it cannot reflect the structure similarity on a wide global scale.} Instead, structure-aware metrics such as SSIM offer a more meaningful assessment of visual quality for some applications. This is further supported by the results in TABLE~\ref{tab:quant_benchmark_results}, where all methods show trivial differences in PSNR, yet SSIM, Homogeneity, \revision{and LPIPS} exhibit clear performance gaps and consistent trends across methods, aligning well with human visual perception.


Interestingly, N2N~\cite{lehtinen2018noise2noise} and DIP~\cite{ulyanov2018deep} are exceptions that show some effectiveness in reducing speckle, though still inferior to traditional methods. This is because DIP does not assume any explicit noise distribution, while N2N assumes independent noise between two observations of the same scene, but does not require the noise within a single image to be independent and pixel-level noise prior. Meanwhile, both are not limited to pixel-level noise modelling. In contrast, other self-supervised methods strongly rely on the assumption of independent noise within a single image, making them ineffective for handling speckle, which exhibits strong spatial and anatomical correlation. However, each has its own limitations: DIP, due to its implicit image prior nature, struggles to distinguish between high-frequency structural details and speckles, making it impossible to determine a reliable early stopping point that removes speckles while preserving anatomical structure. On the other hand, N2N relies on another assumption that noise is zero-mean, which contradicts the nature of speckles. In B-mode \revision{US image}, speckle is typically modelled by non-zero mean distributions such as Nakagami or Rayleigh~\cite{christensen2024systematized}, violating N2N’s assumption and limiting its applicability to ultrasound despeckling.
To intuitively demonstrate the speckle reduction performance, the ground truth and denoised results obtained by different methods are shown in Fig.~\ref{img:comparison_synthetic}. \revision{The inference time is approximately 
$7~ms$, which supports real-time performance, as commercial US machines typically operate at a frame rate of $30$–$60$~Hz for B-mode image generation.}



\subsubsection{Despeckling Performance on Human Carotid Images}
In the absence of ground truth, we used the GLCM homogeneity measure to evaluate the speckle reduction performance on in vivo carotid ultrasound images. As summarized in TABLE~\ref{tab:quant_benchmark_results}, our method achieved the highest homogeneity score (0.725) among all applicable methods, excluding Noise2True and Noise2Noise, which are not applicable in real clinical scenarios. This demonstrates our self-supervised approach's robustness and practical utility in situations where clean reference images are unavailable.

Consistent with the results observed on synthetic data, our method also performs best on the more challenging real-world carotid ultrasound images. Among self-supervised methods, only DIP partially alters image texture but fails to remove speckles effectively. N2N cannot be applied due to its requirement for paired observations, which is infeasible in commonly commercial US devices. Other self-supervised methods show almost no effect on large-scale, granular speckle structures. While traditional methods can partially suppress speckle, they tend to introduce severe textural artefacts and damage fine details. For instance, US image details, such as parallel white lines caused by reverberation artifacts in Fig.~\ref{img:comparison_CCA}, could be used as indicators of structural preservation, but are almost totally degraded by traditional methods.
In contrast, our method achieves a more favourable trade-off, significantly reducing speckle while maintaining anatomical integrity, demonstrating strong potential for real-world clinical applications. Visual comparisons are provided in Fig.~\ref{img:comparison_CCA}.

\begin{figure}[ht!]
\centering
    \includegraphics[width=1\linewidth]{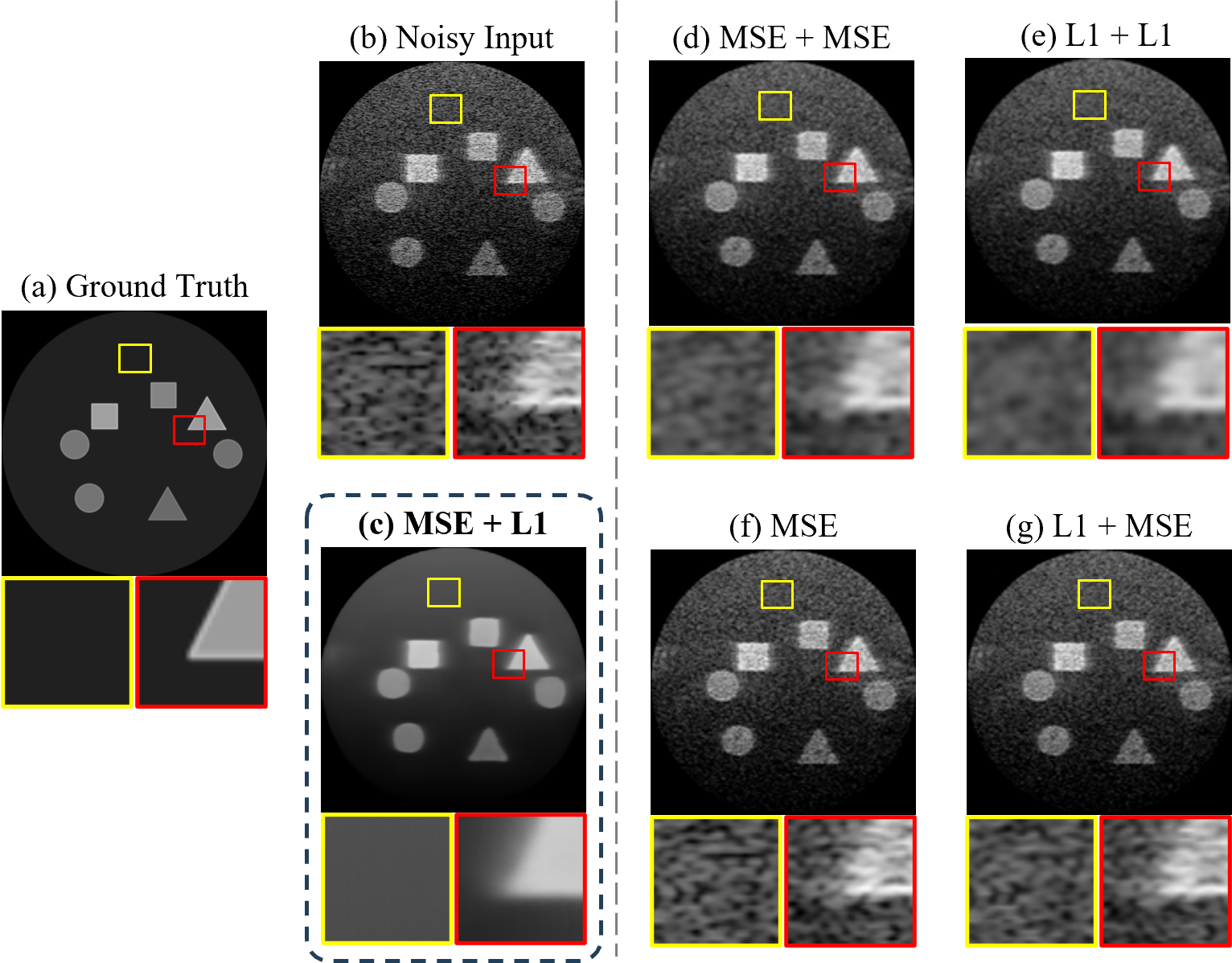}
    \caption{Visual comparison of different loss configurations for training the proposed \revision{Speckle2Self} network. (a) Clean ground truth. (b) Noisy input. (c) \textbf{Our method (MSE + L1)} achieves the best trade-off between speckle suppression and structural fidelity. (d)–(e) Homogeneous losses (L1+L1, MSE+MSE) lead to median-/mean-like smoothing across scales, blurring structure and retaining speckle. (f) MSE-only reconstruction. (g) L1 + MSE enforces overfitting speckle variations and degenerates into single-scale MSE behaviour.}
    \label{fig:ablation_loss}
\end{figure}

\subsection{Ablation Study}

\subsubsection{Effect of Loss Configuration on Despeckling.}
To investigate how different configurations of reconstruction loss $\mathcal{L}_{rec}$ and consistency loss $\mathcal{L}_{cons}$ affect speckle removal performance, we first evaluated five combinations of losses in Eq. (\ref{eq:loss_sum}): 1) MSE for $\mathcal{L}_{rec}$ and L1 for $\mathcal{L}_{cons}$, 2) MSE for both, 3) L1 for both, 4) L1 for $\mathcal{L}_{rec}$ and MSE for $\mathcal{L}_{cons}$, and 5) only use MSE for $\mathcal{L}_{rec}$ without $\mathcal{L}_{cons}$. The quantitative and visual results in TABLE~\ref{tab:ablation} and Fig.~\ref{fig:ablation_loss} (c) demonstrate that the combination of MSE for $\mathcal{L}_{rec}$ with L1 for $\mathcal{L}_{cons}$ (MSE$+$L1) achieves the best performance across all scenarios (S-I, S-II, and S-III). 
Experiments show that, surprisingly, only the combination of (MSE$+$L1) effectively reduces speckle noise, while other combinations lead to near-trivial image reconstruction without a despeckling effect. 
\revision{Since speckle noise typically appears as white grainy patterns in US images~\cite{krissian2005speckle}, the use of MSE (based on a squaring operation) tends to emphasize pixels with higher intensity variations, effectively placing more weight on the grainy noisy areas. This property can help preserve the differences introduced by the MSP operation during reconstruction. In addition, the L1 consistency loss is less sensitive to outliers (each pixel equally contributes), making it better suited for suppressing inconsistent speckles across scales and finding the common invariant tissue structure.}
In contrast:

\begin{itemize}
\item \textbf{MSE reconstruction + MSE consistency:} 
\revision{Replacing the L1 with MSE to enforce consistency between the reconstructed images from two different branches will result in a high attention to recovering the white grainy speckle patterns due to the squaring operation. This can lead to excessive smoothing and substantial boundary distortion, primarily due to the prolonged influence of speckle noise, as illustrated in Fig.~\ref{fig:ablation_loss} (d).}



\item \textbf{L1 reconstruction + L1 consistency:} \revision{Replacing the MSE with L1 for reconstruction will reduce the weight for recovering white grainy speckle patterns.} As a result, it blurs both fine structures and speckle, failing to disentangle them effectively, as shown in Fig.~\ref{fig:ablation_loss} (e). \revision{Compared to Fig.\ref{fig:ablation_loss} (d), Fig.\ref{fig:ablation_loss} (e) shows smoother denoised results with fewer prominent grainy artifacts. This is because the L1 loss equally averages intensity variations across all pixel locations.}



\item \textbf{MSE reconstruction only (no consistency loss):} Without inter-scale consistency constraints, the model overfits each scale individually. This leads to preservation of speckle noise along with structure, as seen in Fig.~\ref{fig:ablation_loss} (f).

\item \textbf{L1 reconstruction + MSE consistency:} This setup performs poorly as the MSE consistency loss is sensitive to outliers, such as speckles. It enforces strict pixel-wise alignment across scales, leading to overfitting of high-frequency speckle noise. The network is driven to minimize all cross-scale differences, effectively collapsing to a single-scale MSE reconstruction that retains speckle artefacts, as observed in Fig.~\ref{fig:ablation_loss} (g).

\end{itemize}

\begin{table}[ht!]
    \scriptsize
    \centering
    \caption{Quantitative results of ablation study in terms of the loss configuration, interpolation scheme, and multi-scale strategy on performance \textbf{(PSNR (dB) / SSIM / Homogeneity)}}
    \resizebox{0.48\textwidth}{!}{
    \begin{tabular}{cccc}
        \toprule
        \textbf{Loss} & \textbf{S-I Data} & \textbf{S-II Data} & \textbf{S-III Data} \\
        \midrule
        \multirow{1}{*}{MSE$+$L1} 
            & \textbf{18.59} / \textbf{0.771} / \textbf{0.803} 
            & \textbf{18.94} / \textbf{0.777} / \textbf{0.792} 
            & \textbf{22.90} / \textbf{0.922} / \textbf{0.876} \\
        \multirow{1}{*}{MSE$+$MSE} 
            & 18.05 / 0.422 / 0.384 
            & 17.72 / 0.399 / 0.383 
            & 21.48 / 0.709 / 0.623 \\
        \multirow{1}{*}{L1$+$L1} 
            & 17.74 / 0.458 / 0.422 
            & 17.74 / 0.439 / 0.425 
            & 21.92 / 0.769 / 0.675 \\
        \multirow{1}{*}{MSE} 
            & 17.80 / 0.316 / 0.315 
            & 17.44 / 0.307 / 0.344 
            & 20.16 / 0.620 / 0.583 \\
        \multirow{1}{*}{L1$+$MSE} 
            & 17.91 / 0.329 / 0.323 
            & 17.47 / 0.319 / 0.345 
            & 20.35 / 0.627 / 0.583 \\
        \midrule
        \midrule
        \textbf{Interp.} & \textbf{S-I Data} & \textbf{S-II Data} & \textbf{S-III Data} \\
        \midrule
        \multirow{1}{*}{Bilinear} 
            & 18.59 / 0.771 / 0.803 
            & \textbf{18.94} / 0.777 / \textbf{0.792} 
            & \textbf{22.90} / \textbf{0.922} / 0.876 \\
        \multirow{1}{*}{Area} 
            & 18.44 / 0.660 / 0.694 
            & 18.63 / 0.657 / 0.678 
            & 22.76 / \textbf{0.922} / 0.872 \\
        \multirow{1}{*}{Bicubic} 
            & \textbf{19.69} / \textbf{0.788} / \textbf{0.830} 
            & 18.70 / \textbf{0.780} / 0.780 
            & 22.30 / 0.918 / \textbf{0.880} \\
        \midrule
        \midrule
        \textbf{Cross-scale} & \textbf{S-I Data} & \textbf{S-II Data} & \textbf{S-III Data} \\
        \midrule
        \multirow{1}{*}{H+M+L} 
            & \textbf{18.59} / \textbf{0.771} / 0.803 
            & \textbf{18.94} / \textbf{0.777} / 0.792 
            & \textbf{22.90} / \textbf{0.922} / \textbf{0.876} \\
        \multirow{1}{*}{H+L} 
            & 18.31 / 0.523 / 0.490 
            & 18.24 / 0.515 / 0.549 
            & 22.22 / 0.786 / 0.720 \\
        \multirow{1}{*}{M+L} 
            & 17.83 / 0.757 / 0.799 
            & 18.71 / 0.776 / 0.830 
            & 22.30 / 0.912 / 0.874 \\
        \multirow{1}{*}{H+M} 
            & 18.37 / 0.500 / 0.485 
            & 18.46 / 0.442 / 0.460 
            & 21.72 / 0.752 / 0.681 \\
        \multirow{1}{*}{OneEncoder} 
            & 17.97 / 0.753 / \textbf{0.824} 
            & 18.38 / 0.764 / \textbf{0.836} 
            & 22.40 / 0.920 / 0.853 \\
        \bottomrule
    \end{tabular}
    }
    \label{tab:ablation}
\end{table}

\subsubsection{Effect of Interpolation Scheme on Despeckling.}
To introduce distinctive variations in speckle noise, we applied the MSP operation $\mathcal{P}_{k}(\cdot)$ at scales $ s_k \in \{1.0, 0.5, 0.25\} $ individually. During the downsampling–upsampling, the choice of interpolation method may impact the final performance. To investigate this, we evaluated three standard interpolation methods: 1) bilinear, 2) area, and 3) bicubic interpolation. As shown in TABLE~\ref{tab:ablation}, both bilinear and bicubic interpolation consistently demonstrated strong performance across all scenarios, effectively enhancing the model’s despeckling capability. 
In contrast, pixel area interpolation underperformed in speckle reduction. This may be attributed to pixel area interpolation's tendency to preserve spatial coherence of speckle patterns by averaging pixel values over an area, which limits its effectiveness in disrupting speckle structures compared to bilinear and bicubic methods.

\par
\begin{figure}[ht!]
\centering
    \includegraphics[width=1\linewidth]{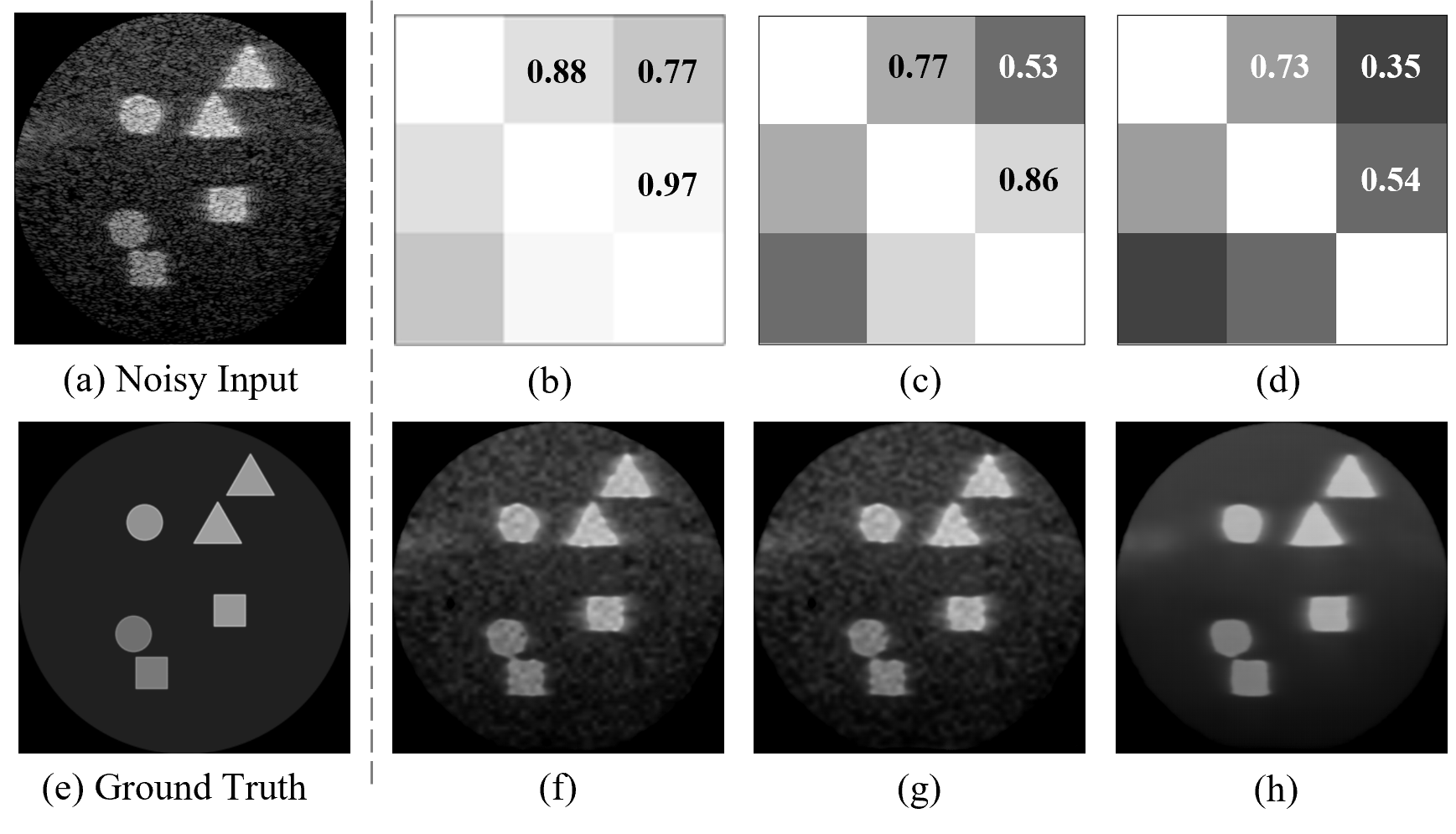}
    \caption{\revision{Denoising performance comparison between low-pass filters and the proposed MSP operation as preprocessing. (a) Noisy input image. (b) and (c) Correlation matrices computed among the high-frequency component of three images generated using Gaussian smoothing filters with two sets of kernel sizes: LPF$^1 = \{1, 3, 5\}$ and LPF$^2 = \{1, 5, 9\}$, respectively. (d) Correlation matrix computed using the proposed MSP operation with scaling factors $s_k \in \{1.0, 0.5, 0.25\}$. (e) Clean target image. (f), (g) and (h) Denoised results using LPF$^1$, LPF$^2$, and the proposed MSP, respectively.}}
    \label{fig:fig_low_pass}
\end{figure}

\revision{
The low-pass filter (LPF) can be used to maintain the low-frequency information while removing the high-frequency component. So LPF can also introduce variations in the high-frequency feature by using different parameters. To investigate the effectiveness of LPF in comparison with MSP operation, we further conduct the experiments using two sets of kernel sizes (1/3/5 and 1/5/9, respectively) in the standard Gaussian smooth filter. The results has on a representative sample image are depicted in Fig.~\ref{fig:fig_low_pass}. }

\par
\revision{
It can be seen from Fig.~\ref{fig:fig_low_pass} (b-d) that the proposed MPS can result in more significant variation in the high-frequency component of the input images than using LPF. In addition, it is worth noting that the speckle was not effectively removed when the LPFs were used. Although LPF$^2$ employs larger kernel sizes than LPF$^1$, it does not lead to a visually noticeable improvement in speckle denoising performance. This may be attributed to the fact that the variations induced by LPFs with different kernel sizes are not independent—for instance, the high-frequency components removed by a filter with a higher cutoff frequency are also included in those removed by a filter with a lower cutoff frequency.}

\noindent
\subsubsection{Effect of Multi-Scale Strategy on Despeckling.}

We further evaluated different model configurations to understand the impact of scale combinations and encoder design:
(1) H+M+L: Three encoders for high, medium, and low resolutions;
(2) H+L: Two encoders for high and low resolutions;
(3) M+L: Two encoders for medium and low resolutions;
(4) H+M: Two encoders for high and medium resolutions;
(5) OneEncoder: A single shared encoder for all scales.
As summarized in TABLE~\ref{tab:ablation}, the H+M+L configuration consistently achieved the best performance across all datasets, confirming that combining inputs across diverse scales facilitates more effective disentanglement of shared structure and scale-sensitive speckle. Among dual-scale variants, M+L outperformed the others, likely due to its moderate resolution gap that introduces sufficient variation in speckle while preserving cross-scale structural alignment. In contrast, H+M provided limited variation, while H+L introduced too much, making inter-scale learning less effective. Interestingly, the OneEncoder setup achieved the second-best results. This suggests that even with a shared encoder, the presence of multi-scale inputs alone allows the network to benefit from cross-scale speckle diversity, though not as effectively as when scale-specific encoders are used.

\subsection{Generalization Capability}
To evaluate the generalisation capability of our method, we conducted cross-configuration experiments on simulated data and cross-device validation on in vivo carotid ultrasound images acquired from different US systems.

\begin{figure}[ht!]
    \small
    \centering
    \setlength{\tabcolsep}{1.5pt} 
    \renewcommand{\arraystretch}{1} 
    \begin{tabular}{ccccc}
        \textbf{\qquad \qquad 3.75 MHz} & 
        \textbf{5.63 MHz} & 
        \textbf{9.38 MHz} & 
        \multirow{2}{*}[-8mm]{\textbf{Clean Target}} \\

        \multirow{1}{*}[8mm]{\textbf{Noised\quad}}
        
        \includegraphics[width=0.10\textwidth]{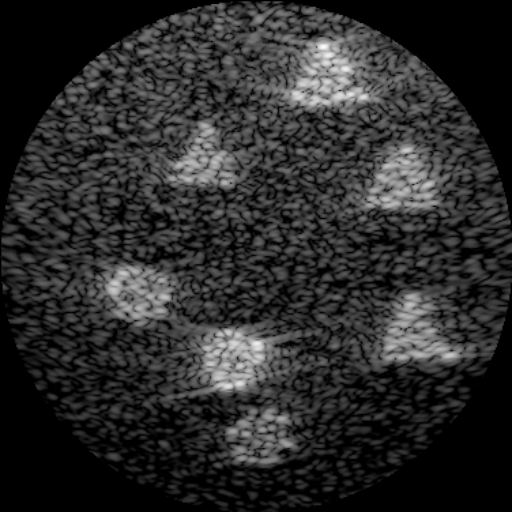} &   
        \includegraphics[width=0.10\textwidth]{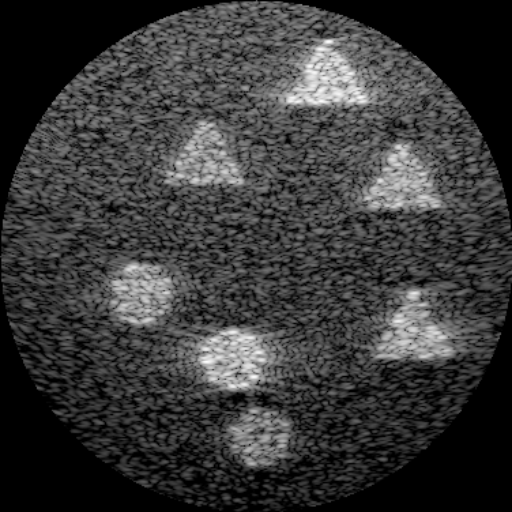} & 
        \includegraphics[width=0.10\textwidth]{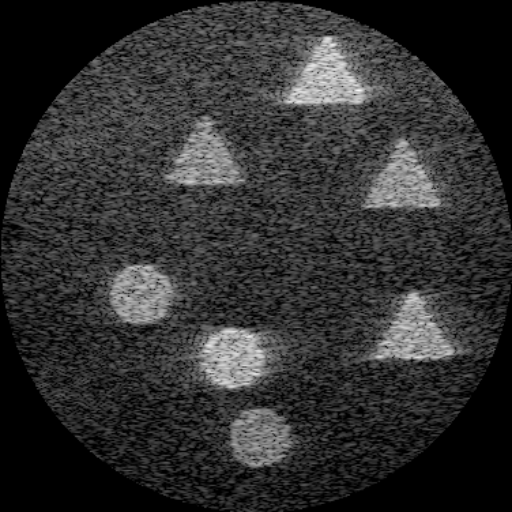} & 
        
        \multirow{2}{*}[5mm]{\includegraphics[width=0.10\textwidth]{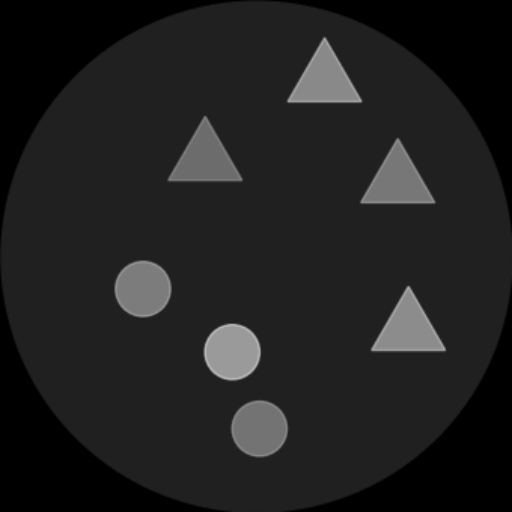}} \\

        
        \multirow{1}{*}[8mm]{\textbf{Denoised}}
        \begin{overpic}[width=0.10\textwidth]{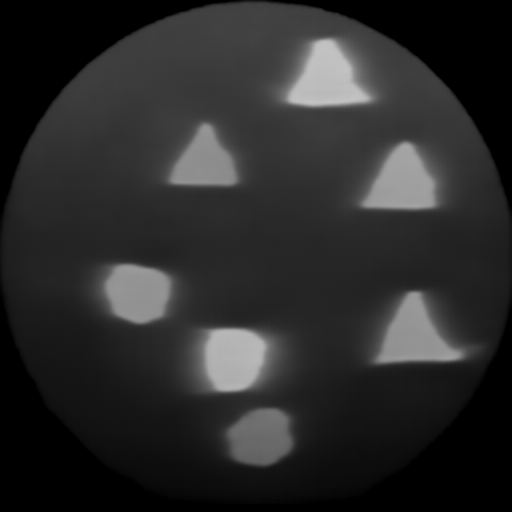} 
        \put(0, 10){\textbf{\textcolor{white}{\scriptsize SSIM: 0.808}}}
        \end{overpic}&
        \begin{overpic}[width=0.10\textwidth]{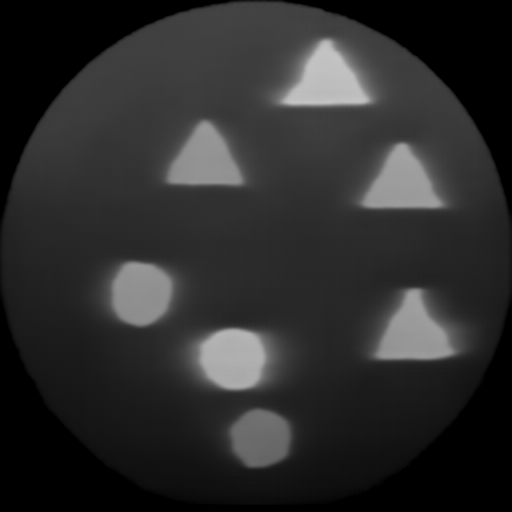} 
        \put(0, 10){\textbf{\textcolor{white}{\scriptsize SSIM: 0.819}}}
        \end{overpic}&
        \begin{overpic}[width=0.10\textwidth]{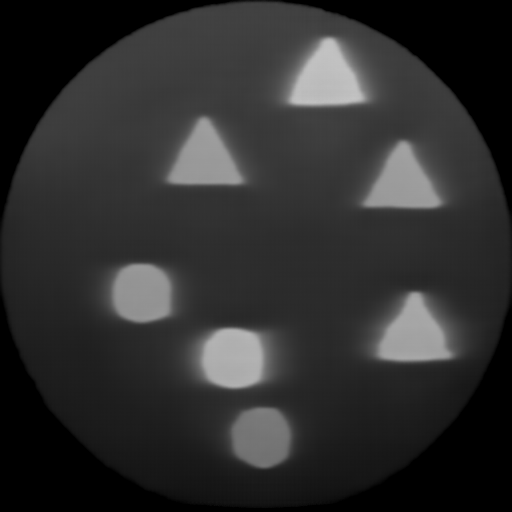} 
        \put(0, 10){\textbf{\textcolor{white}{\scriptsize SSIM: 0.832}}}
        \end{overpic}& \\

    \end{tabular}
    \caption{Cross-configuration testing results show that higher central frequencies generate finer image textures, improving SSIM metrics.
    }
    \label{fig:cross_conf}
\end{figure}

\subsubsection{Cross-Configuration Testing on Synthetic Data} 
\label{sec:cross_conf}
We generated three additional versions of US images of the same scenario by varying simulator configurations. All settings used the same L12-3V virtual probe, but with different central frequencies $f_{c}$ of 3.75, 5.63, and 9.38 MHz, respectively, producing unseen speckle patterns for the model originally trained at $f_{c}$ of 7.50 MHz. The detailed parameters are shown in TABLE~\ref{tab:para_sim}. 
Without any fine-tuning, the model demonstrates strong generalization capability with unseen data, achieving SSIM scores of 0.808, 0.819, and 0.832 on the new configurations. Interestingly, despeckling performance improves with increasing frequency, as higher-frequency US images preserve finer structural details, resulting in more distinguishable speckle patterns after the \revision{MSP} operation. A visual comparison is shown in Fig.~\ref{fig:cross_conf}.

\begin{figure}[ht!]
\centering
    \includegraphics[width=1\linewidth]{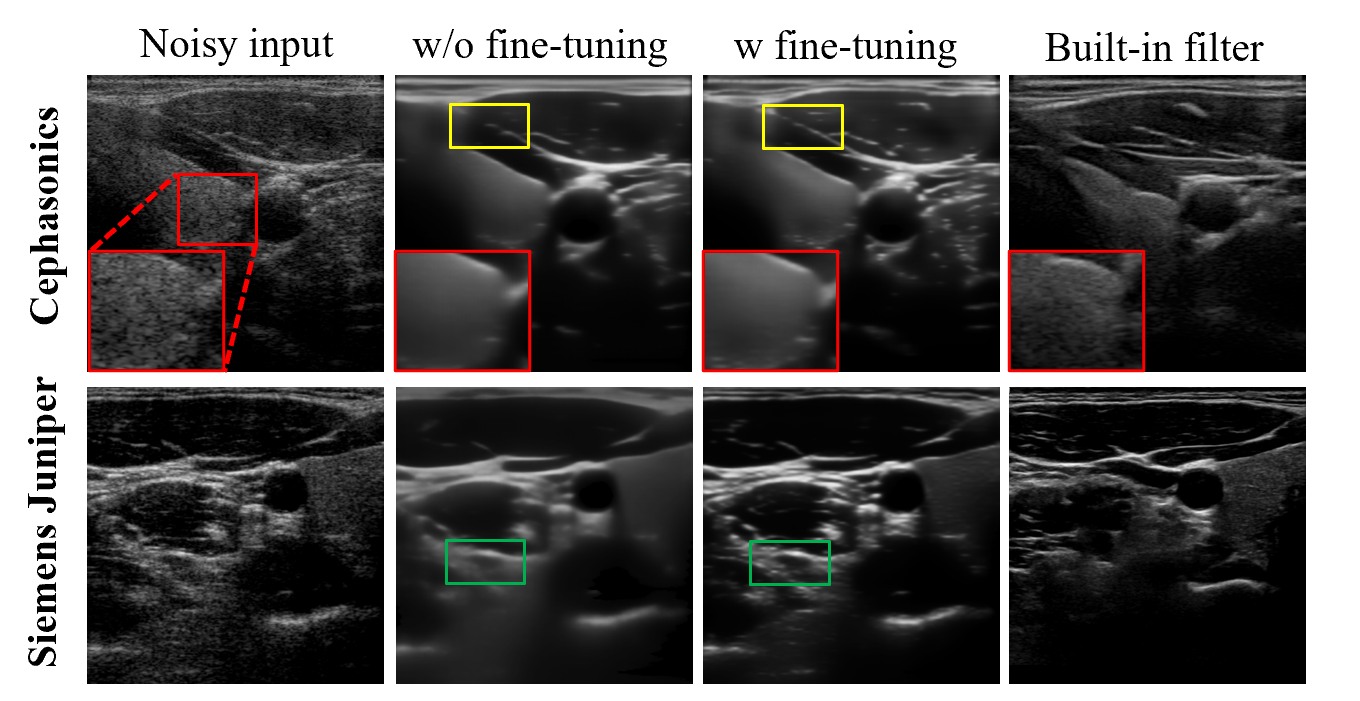}
    \caption{Qualitative comparison of denoising performance on images obtained from two unseen carotid US images obtained using two different US devices. The top row depicts the US image obtained from the Cephasonics US machine, while the bottom one is from the Siemens Juniper machine. The red rectangle shows imagined details, and the yellow/green rectangle annotate the better structural preservation after fine-turning.}
    \label{fig:device_compare}
\end{figure}

\begin{figure*}[ht!]
    \small
    \centering
    \setlength{\tabcolsep}{1.5pt} 
    \renewcommand{\arraystretch}{1} 
    \begin{tabular}{c}
        \begin{tabular}{ccccc}
                \textbf{Noisy Input} & 
                \textbf{BM3D} & 
                \textbf{NLM} & 
                \textbf{OBNLM} & 
                \textbf{SRAD}
                \\
        
                \includegraphics[width=0.14\textwidth]{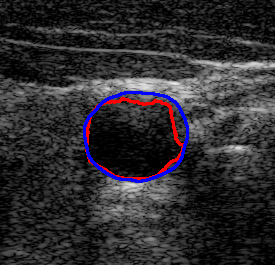} & 
                \includegraphics[width=0.14\textwidth]{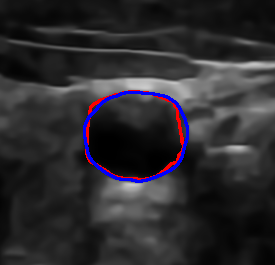} & 
                \includegraphics[width=0.14\textwidth]{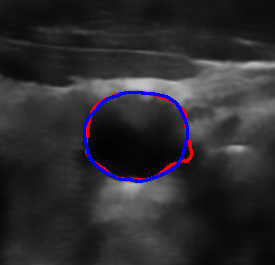} & 
                \includegraphics[width=0.14\textwidth]{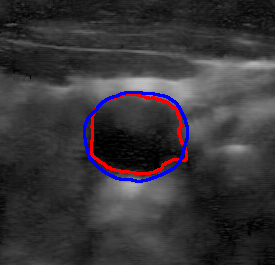} & 
                \includegraphics[width=0.14\textwidth]{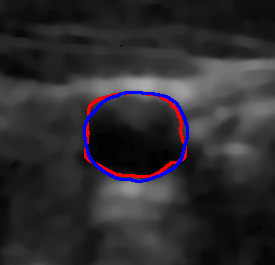} \\

        \end{tabular}\\

        \begin{tabular}{cccccc}
                \textbf{Our Method} & 
                \textbf{N2V} & 
                \textbf{N2S} & 
                \textbf{DIP(*)} &
                \textbf{ZS-N2N(*)} &
                \textbf{Neigh2Neigh}\\  
        
                \includegraphics[width=0.14\textwidth]{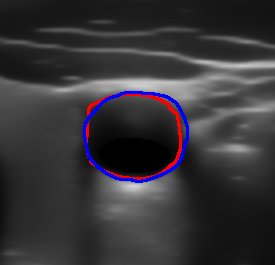} & 
                \includegraphics[width=0.14\textwidth]{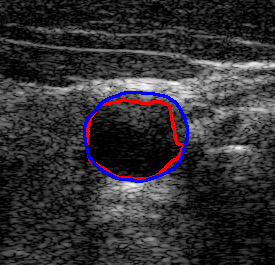} & 
                \includegraphics[width=0.14\textwidth]{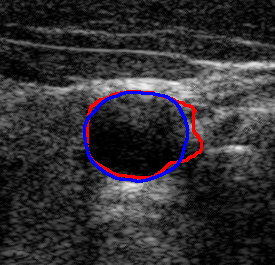} & 
                \includegraphics[width=0.14\textwidth]{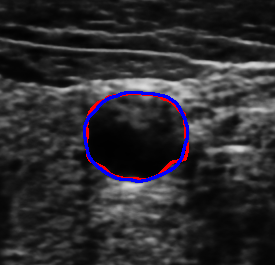} & 
                \includegraphics[width=0.14\textwidth]{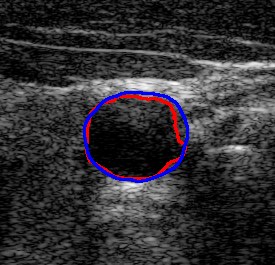} &
                \includegraphics[width=0.14\textwidth]{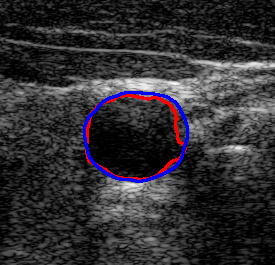} \\
    
        \end{tabular}
    \end{tabular}
    \caption{Comparison of various speckle reduction methods in enhancing carotid ultrasound segmentation accuracy: Blue curves indicate the ground truth boundaries delineated by an experienced ultrasound physician, while red curves represent the segmentation results generated by SAM2 on the original noisy US images and those processed by different despeckling methods.}
    \label{fig:downstream_CCA}
\end{figure*}

\subsubsection{Cross-Device Testing on Carotid US Images}
\label{sec:cross_device}
To further assess the model’s generalization capability on real US data, we collected two additional smaller datasets from different US machines (each has 180 frames), as summarized in TABLE~\ref{tab:para_device}. For cross-device generalization, we first applied the model—trained solely on Clarius data directly to noisy images from these unseen devices, without any fine-tuning. \revision{As shown in Fig.~\ref{fig:device_compare}}, the model still can effectively suppress speckle on these unseen domain images, demonstrating strong zero-shot generalization. 
As a qualitative comparison, we also included denoised results from the commercial built-in filters of the respective ultrasound devices. Compared to these built-in denoisers, our method achieves superior speckle suppression while better preserving structural detail. This is particularly evident in homogeneous regions such as the thyroid, as illustrated in Fig.~\ref{fig:device_compare}.
To further optimize performance for each unseen domain, we then performed lightweight fine-tuning on each dataset for 1000 epochs (approximately 30 minutes). The results show that the model achieves improved denoising quality even with minimal adaptation. As shown in Fig.~\ref{fig:device_compare}, improvements in structural preservation—especially in the thyroid region—can be observed after short fine-tuning, as highlighted by the yellow rectangle for Cephasonics and the green rectangle for Siemens. These results further validate the adaptability and effectiveness of the proposed method across domains.

\subsection{Downstream Task on Ultrasound Image Segmentation}
\par
\revision{
To showcase the potential of the proposed denoising method quantitatively for ultrasound image understanding, we evaluate its impact on carotid ultrasound segmentation as a downstream task~\cite{chen2024towards}. In this task, we use $120$ unseen noisy carotid images collected from the Clarius device. Then, unsupervised Segment Anything 2 (SAM2)~\cite{ravi2024sam} and the rule-based Flood Fill algorithm are used to extract the carotid artery from noisy and denoised images processed by various despeckling algorithms. In both cases, we place the seed point at the geometric center of the carotid lumen in each frame.}

\par
\revision{
The quantitative result in terms of both IoU and AUC has been summarized in Table~\ref{tab:quant_downstream}. To provide an intuitive visualization of the comparison, the results of SAM2 on a representative image and its denoised variants are presented in Fig.~\ref{fig:downstream_CCA}. It can be seen from Table~\ref{tab:quant_downstream} that the overall results using SAM2 are significantly improved in comparison to the Folld fill algorithm. This is primarily because of the strong generalization capability of the foundation model itself. Although the results produced by SAM2 across different images are comparable, the outcomes on images denoised by our method consistently demonstrate superior performance. The AUC reaches the highest score among all methods, while the IoU achieves the second-best performance. This performance advantage becomes more pronounced when using traditional region-growing algorithms such as the Flood Fill method. Significant improvements in both AUC ($0.816$ vs $0.589$) and IoU ($0.497$ vs $0.175$) are observed when applying the proposed Speckle2Self denoising approach.
}

\begin{table}[ht!]
    \centering
    \caption{\revision{Quantitative comparison of different methods for US speckle reduction in terms of IoU and AUC, based on both SAM2- and Flood Fill–based segmentation. The best-performing method for each metric is highlighted in bold.}}
    \resizebox{0.45\textwidth}{!}{
    \begin{tabular}{lcc}
        \toprule
        \multirow{2}{*}{\textbf{Method}} & \textbf{SAM2} & \textbf{Flood Fill} \\
        & IoU($\uparrow$)/AUC($\uparrow$) & IoU($\uparrow$)/AUC($\uparrow$) \\
        \midrule
        Without Denoising & 0.868 / 0.945 & 0.175 / 0.589 \\
        \midrule
        SRAD~\cite{yu2002speckle} & 0.884 / 0.950 & 0.208 / 0.604 \\
        NLM~\cite{buades2005non} & \textbf{0.918} / 0.982 & 0.273 / 0.637 \\
        BM3D~\cite{dabov2007image} & 0.880 / 0.982 & 0.271 / 0.636 \\
        OBNLM~\cite{coupe2009nonlocal} & 0.811 / 0.907 & 0.028 / 0.514 \\
        \midrule
        DIP(*)~\cite{ulyanov2018deep} & 0.853 / 0.950 & 0.304 / 0.653 \\
        N2V~\cite{krull2019noise2void} & 0.835 / 0.927 & 0.170 / 0.586 \\
        N2S(*)~\cite{batson2019noise2self} & 0.819 / 0.930 & 0.166 / 0.583 \\
        N2S~\cite{batson2019noise2self} & 0.819 / 0.930 & 0.166 / 0.583 \\
        Neighbor2Neighbor~\cite{huang2021neighbor2neighbor} & 0.895 / 0.960 & 0.170 / 0.586 \\
        ZS-N2N(*)~\cite{mansour2023zero} & 0.895 / 0.960 & 0.209 / 0.605 \\
        Speckle2Self & 0.914 / \textbf{0.985} & \textbf{0.497} / \textbf{0.816} \\
        \bottomrule
    \end{tabular}
    }
    \label{tab:quant_downstream}
\end{table}

\section{\revision{Discusssion}}

\revision{
In this work, we proposed a novel self-supervised despeckling framework for B-mode US images that requires neither clean reference data nor multiple observations of the same scene. Experimental results demonstrate that the proposed method consistently outperforms recent self-supervised approaches and traditional filtering techniques on both synthetic datasets and in vivo carotid scans. Furthermore, the method exhibits strong generalization to unseen domains, including variations in probe settings and US imaging devices. Despite its advantages, the approach has certain limitations. Due to the down-sampling involved in the MSP operation, the resulting denoised images exhibit slight blurring along tissue boundaries, which may be suboptimal for diagnostic purposes—particularly for clinicians accustomed to interpreting images with speckle patterns. As noted in prior work~\cite{demir2025diffdenoise, nagare2021bias}, the use of self-supervised learning frameworks and MSE loss functions may also result in relatively smooth outputs. Nonetheless, our collaborating clinicians have confirmed that the denoised images are promising for specific applications such as anatomical geometry measurement and annotation tasks, where reduced speckle will enhance clarity and precision.}

\par
In addition, the proposed method does not require explicitly specifying the statistical distribution of speckle, such as assuming Gaussian or other models, as adopted in some prior works. Instead, it relies on a classical low-rank prior, wherein anatomical structures in US are assumed to exhibit low-rank characteristics, while speckle components are not. Although, as discussed in Section~\ref{subsec:low_rank}, this assumption has been widely applied in US signal processing, it remains an approximation rather than a physically rigorous model. Nevertheless, it has demonstrated practical effectiveness across diverse imaging conditions. In future work, we aim to incorporate more physics-informed priors of speckle formation, \revisionSecond{such as its depth-related spatial characteristics discussed in~\cite{goudarzi2023deep}, to further improve the interpretability and robustness.}


\revision{
Our approach also operates on B-mode US images rather than RF signals. While this enhances compatibility with clinical workflows, it may limit access to the richer information in RF data, which preserves phase and offers a higher dynamic range. 
However, leveraging RF data also introduces additional complexity in acquisition and preprocessing. Effectively integrating RF signals into the learning framework remains an important direction for future work, as explored in prior studies (e.g.,~\cite{stevens2024dehazing,sharifzadeh2024mitigating}). By leveraging the rich and dynamic information in RF data, this integration may help mitigate the over-smoothed tissue boundaries observed in the current study.
}


It is also worth noting that, although speckle often introduces severe visual quality degradation in B-mode US. However, as discussed in Section~\ref{sec:us_property}, US speckle arises not from a random process, but a deterministic process~\cite{szabo2013diagnostic, hoskins2019diagnostic}. Given the same scene and probe configurations, the speckle pattern remains almost identical, making it a valuable signal for motion analysis~\cite{jiang2021deformation, jiang2023defcor}, known as speckle tracking. This technique underlies a wide range of applications, such as elastography~\cite{curiale2016influence, wen2023convolutional}, blood flow estimation~\cite{nyrnes2020blood}, and echocardiography~\cite{lu2023ultrafast}.
While the proposed method focuses on speckle suppression to enhance image quality, its ability to separate speckle from tissue structures may also facilitate downstream analysis of speckle patterns. Thus, the framework has potential not only for despeckling but also for speckle-based applications.
%

\section{Conclusion}
We present Speckle2Self, a novel self-supervised framework for US speckle reduction that operates using only single noisy B-mode images. By applying the proposed MSP operation, \revision{Speckle2Self} generates variant inputs that preserve shared anatomical structure while inducing scale-dependent speckle variation. This enables the model to learn cross-scale consistency and implicitly disentangle clean tissue from speckle, without requiring clean references or paired observations. 
Extensive evaluations on both realistic simulated and in vivo carotid ultrasound data demonstrate that \revision{Speckle2Self} consistently outperforms traditional filters and recent self-supervised learning-based methods. Moreover, \revision{Speckle2Self} exhibits strong generalization across domains—including unseen probe frequencies and ultrasound devices—highlighting its robustness and clinical applicability. This denied result will enhance image consistency and thereby boost the development of reproducible computer-aided diagnosis, such as the emerging topic of VLM-based automated US examination report generation~\cite{li2024ultrasound,li2025semantic}. We believe this data-efficient, architecture-light, and hardware-independent framework holds great promise for scalable deployment in portable and point-of-care ultrasound systems, especially in low-resource settings. This advance will lay the ground for affordable, universal healthcare intervention by enhancing the low-cost US imaging quality.

\section*{Declaration of Competing Interest}
\label{competinginterest}

The authors report no conflicts of interest.


\section*{Acknowledgments}
The authors want to thank the clinical partner, Dr. Angelos Karlas, for his insightful discussion throughout the course of this study, and acknowledge the Editors and all reviewers for their time and implicit contributions to the improvement of the article's thoroughness, readability, and clarity. This work was supported in part by Brainlab, Germany; as well as in part by the Multi-Scale Medical Robotics Center, AIR@InnoHK, Hong Kong; and in part by the SINO-German Mobility Project under Grant M0221.

\bibliographystyle{model2-names.bst}\biboptions{authoryear}
\bibliography{references}

\begin{thebibliography}{91}
\expandafter\ifx\csname natexlab\endcsname\relax\def\natexlab#1{#1}\fi
\providecommand{\url}[1]{\texttt{#1}}
\providecommand{\href}[2]{#2}
\providecommand{\path}[1]{#1}
\providecommand{\DOIprefix}{doi:}
\providecommand{\ArXivprefix}{arXiv:}
\providecommand{\URLprefix}{URL: }
\providecommand{\Pubmedprefix}{pmid:}
\providecommand{\doi}[1]{\href{http://dx.doi.org/#1}{\path{#1}}}
\providecommand{\Pubmed}[1]{\href{pmid:#1}{\path{#1}}}
\providecommand{\bibinfo}[2]{#2}
\ifx\xfnm\relax \def\xfnm[#1]{\unskip,\space#1}\fi
\bibitem[{Aja-Fern{\'a}ndez and Alberola-L{\'o}pez(2006)}]{aja2006estimation}
\bibinfo{author}{Aja-Fern{\'a}ndez, S.}, \bibinfo{author}{Alberola-L{\'o}pez, C.}, \bibinfo{year}{2006}.
\newblock \bibinfo{title}{On the estimation of the coefficient of variation for anisotropic diffusion speckle filtering}.
\newblock \bibinfo{journal}{IEEE Transactions on Image Processing} \bibinfo{volume}{15}, \bibinfo{pages}{2694--2701}.
\bibitem[{Asgariandehkordi et~al.(2025)Asgariandehkordi, Sharifzadeh and Rivaz}]{asgariandehkordi2025lightweight}
\bibinfo{author}{Asgariandehkordi, H.}, \bibinfo{author}{Sharifzadeh, M.}, \bibinfo{author}{Rivaz, H.}, \bibinfo{year}{2025}.
\newblock \bibinfo{title}{Lightweight physics-informed zero-shot ultrasound plane wave denoising}.
\newblock \bibinfo{journal}{arXiv preprint arXiv:2506.21499} .
\bibitem[{Batson and Royer(2019)}]{batson2019noise2self}
\bibinfo{author}{Batson, J.}, \bibinfo{author}{Royer, L.}, \bibinfo{year}{2019}.
\newblock \bibinfo{title}{Noise2self: Blind denoising by self-supervision}, in: \bibinfo{booktitle}{International Conference on Machine Learning}, \bibinfo{organization}{PMLR}. pp. \bibinfo{pages}{524--533}.
\bibitem[{Bi et~al.(2025)Bi, Huang, Clarenbach, Ghotbi, Karlas, Navab and Jiang}]{bi2025synomaly}
\bibinfo{author}{Bi, Y.}, \bibinfo{author}{Huang, L.}, \bibinfo{author}{Clarenbach, R.}, \bibinfo{author}{Ghotbi, R.}, \bibinfo{author}{Karlas, A.}, \bibinfo{author}{Navab, N.}, \bibinfo{author}{Jiang, Z.}, \bibinfo{year}{2025}.
\newblock \bibinfo{title}{Synomaly noise and multi-stage diffusion: A novel approach for unsupervised anomaly detection in medical images}.
\newblock \bibinfo{journal}{Medical Image Analysis} , \bibinfo{pages}{103737}.
\bibitem[{Bi et~al.(2023a)Bi, Jiang, Clarenbach, Ghotbi, Karlas and Navab}]{bi2023mi}
\bibinfo{author}{Bi, Y.}, \bibinfo{author}{Jiang, Z.}, \bibinfo{author}{Clarenbach, R.}, \bibinfo{author}{Ghotbi, R.}, \bibinfo{author}{Karlas, A.}, \bibinfo{author}{Navab, N.}, \bibinfo{year}{2023}a.
\newblock \bibinfo{title}{Mi-segnet: Mutual information-based us segmentation for unseen domain generalization}, in: \bibinfo{booktitle}{International Conference on Medical Image Computing and Computer-Assisted Intervention}, \bibinfo{organization}{Springer}. pp. \bibinfo{pages}{130--140}.
\bibitem[{Bi et~al.(2023b)Bi, Jiang, Duelmer, Huang and Navab}]{bi2024machine}
\bibinfo{author}{Bi, Y.}, \bibinfo{author}{Jiang, Z.}, \bibinfo{author}{Duelmer, F.}, \bibinfo{author}{Huang, D.}, \bibinfo{author}{Navab, N.}, \bibinfo{year}{2023}b.
\newblock \bibinfo{title}{Machine learning in robotic ultrasound imaging: Challenges and perspectives}.
\newblock \bibinfo{journal}{Annual Review of Control, Robotics, and Autonomous Systems} \bibinfo{volume}{7}.
\bibitem[{Buades et~al.(2005)Buades, Coll and Morel}]{buades2005non}
\bibinfo{author}{Buades, A.}, \bibinfo{author}{Coll, B.}, \bibinfo{author}{Morel, J.M.}, \bibinfo{year}{2005}.
\newblock \bibinfo{title}{A non-local algorithm for image denoising}, in: \bibinfo{booktitle}{2005 IEEE computer society conference on computer vision and pattern recognition (CVPR'05)}, \bibinfo{organization}{Ieee}. pp. \bibinfo{pages}{60--65}.
\bibitem[{Calis et~al.(2025)Calis, Mischi, van~der Veen and Hunyadi}]{calis2025speckle}
\bibinfo{author}{Calis, M.}, \bibinfo{author}{Mischi, M.}, \bibinfo{author}{van~der Veen, A.J.}, \bibinfo{author}{Hunyadi, B.}, \bibinfo{year}{2025}.
\newblock \bibinfo{title}{Speckle denoising of dynamic contrast-enhanced ultrasound using low-rank tensor decomposition}.
\newblock \bibinfo{journal}{IEEE Transactions on Medical Imaging} .
\bibitem[{Chen et~al.(2024)Chen, Chen, Song, Xiong, Yuille, Wei and Zhou}]{chen2024towards}
\bibinfo{author}{Chen, Q.}, \bibinfo{author}{Chen, X.}, \bibinfo{author}{Song, H.}, \bibinfo{author}{Xiong, Z.}, \bibinfo{author}{Yuille, A.}, \bibinfo{author}{Wei, C.}, \bibinfo{author}{Zhou, Z.}, \bibinfo{year}{2024}.
\newblock \bibinfo{title}{Towards generalizable tumor synthesis}, in: \bibinfo{booktitle}{Proceedings of the IEEE/CVF conference on computer vision and pattern recognition}, pp. \bibinfo{pages}{11147--11158}.
\bibitem[{Chen et~al.(2010)Chen, Montesinos, Sun et~al.}]{chen2010ramp}
\bibinfo{author}{Chen, Q.}, \bibinfo{author}{Montesinos, P.}, \bibinfo{author}{Sun, Q.S.}, et~al., \bibinfo{year}{2010}.
\newblock \bibinfo{title}{Ramp preserving perona--malik model}.
\newblock \bibinfo{journal}{Signal Processing} \bibinfo{volume}{90}, \bibinfo{pages}{1963--1975}.
\bibitem[{Chen et~al.(2025)Chen, Fang, Li, Huang and Luo}]{chen2025ultrafast}
\bibinfo{author}{Chen, Y.}, \bibinfo{author}{Fang, B.}, \bibinfo{author}{Li, H.}, \bibinfo{author}{Huang, L.}, \bibinfo{author}{Luo, J.}, \bibinfo{year}{2025}.
\newblock \bibinfo{title}{Ultrafast online clutter filtering for ultrasound microvascular imaging}.
\newblock \bibinfo{journal}{IEEE Transactions on Medical Imaging} .
\bibitem[{Cho et~al.(2024)Cho, Park, Kang and Yoo}]{cho2024deep}
\bibinfo{author}{Cho, H.}, \bibinfo{author}{Park, S.}, \bibinfo{author}{Kang, J.}, \bibinfo{author}{Yoo, Y.}, \bibinfo{year}{2024}.
\newblock \bibinfo{title}{Deep coherence learning: An unsupervised deep beamformer for high quality single plane wave imaging in medical ultrasound}.
\newblock \bibinfo{journal}{Ultrasonics} \bibinfo{volume}{143}, \bibinfo{pages}{107408}.
\bibitem[{Christensen et~al.(2024)Christensen, Rosado-Mendez and Hall}]{christensen2024systematized}
\bibinfo{author}{Christensen, A.M.}, \bibinfo{author}{Rosado-Mendez, I.M.}, \bibinfo{author}{Hall, T.J.}, \bibinfo{year}{2024}.
\newblock \bibinfo{title}{A systematized review of quantitative ultrasound based on first-order speckle statistics}.
\newblock \bibinfo{journal}{IEEE Transactions on Ultrasonics, Ferroelectrics, and Frequency Control} .
\bibitem[{Cigier et~al.(2022)Cigier, Varray and Garcia}]{cigier2022simus}
\bibinfo{author}{Cigier, A.}, \bibinfo{author}{Varray, F.}, \bibinfo{author}{Garcia, D.}, \bibinfo{year}{2022}.
\newblock \bibinfo{title}{Simus: an open-source simulator for medical ultrasound imaging. part ii: comparison with four simulators}.
\newblock \bibinfo{journal}{Computer Methods and Programs in Biomedicine} \bibinfo{volume}{220}, \bibinfo{pages}{106774}.
\bibitem[{Coup{\'e} et~al.(2009)Coup{\'e}, Hellier, Kervrann and Barillot}]{coupe2009nonlocal}
\bibinfo{author}{Coup{\'e}, P.}, \bibinfo{author}{Hellier, P.}, \bibinfo{author}{Kervrann, C.}, \bibinfo{author}{Barillot, C.}, \bibinfo{year}{2009}.
\newblock \bibinfo{title}{Nonlocal means-based speckle filtering for ultrasound images}.
\newblock \bibinfo{journal}{IEEE transactions on image processing} \bibinfo{volume}{18}, \bibinfo{pages}{2221--2229}.
\bibitem[{Curiale et~al.(2016)Curiale, Vegas-S{\'a}nchez-Ferrero and Aja-Fern{\'a}ndez}]{curiale2016influence}
\bibinfo{author}{Curiale, A.H.}, \bibinfo{author}{Vegas-S{\'a}nchez-Ferrero, G.}, \bibinfo{author}{Aja-Fern{\'a}ndez, S.}, \bibinfo{year}{2016}.
\newblock \bibinfo{title}{Influence of ultrasound speckle tracking strategies for motion and strain estimation}.
\newblock \bibinfo{journal}{Medical image analysis} \bibinfo{volume}{32}, \bibinfo{pages}{184--200}.
\bibitem[{Dabov et~al.(2007)Dabov, Foi, Katkovnik and Egiazarian}]{dabov2007image}
\bibinfo{author}{Dabov, K.}, \bibinfo{author}{Foi, A.}, \bibinfo{author}{Katkovnik, V.}, \bibinfo{author}{Egiazarian, K.}, \bibinfo{year}{2007}.
\newblock \bibinfo{title}{Image denoising by sparse 3-d transform-domain collaborative filtering}.
\newblock \bibinfo{journal}{IEEE Transactions on image processing} \bibinfo{volume}{16}, \bibinfo{pages}{2080--2095}.
\bibitem[{Dantas et~al.(2005)Dantas, Costa and Leeman}]{dantas2005ultrasound}
\bibinfo{author}{Dantas, R.G.}, \bibinfo{author}{Costa, E.T.}, \bibinfo{author}{Leeman, S.}, \bibinfo{year}{2005}.
\newblock \bibinfo{title}{Ultrasound speckle and equivalent scatterers}.
\newblock \bibinfo{journal}{Ultrasonics} \bibinfo{volume}{43}, \bibinfo{pages}{405--420}.
\bibitem[{Demen{\'e} et~al.(2015)Demen{\'e}, Deffieux, Pernot, Osmanski, Biran, Gennisson, Sieu, Bergel, Franqui, Correas et~al.}]{demene2015spatiotemporal}
\bibinfo{author}{Demen{\'e}, C.}, \bibinfo{author}{Deffieux, T.}, \bibinfo{author}{Pernot, M.}, \bibinfo{author}{Osmanski, B.F.}, \bibinfo{author}{Biran, V.}, \bibinfo{author}{Gennisson, J.L.}, \bibinfo{author}{Sieu, L.A.}, \bibinfo{author}{Bergel, A.}, \bibinfo{author}{Franqui, S.}, \bibinfo{author}{Correas, J.M.}, et~al., \bibinfo{year}{2015}.
\newblock \bibinfo{title}{Spatiotemporal clutter filtering of ultrafast ultrasound data highly increases doppler and fultrasound sensitivity}.
\newblock \bibinfo{journal}{IEEE transactions on medical imaging} \bibinfo{volume}{34}, \bibinfo{pages}{2271--2285}.
\bibitem[{Demir et~al.(2025)Demir, Liu, Chen, Chen, Zhao, Mailhe, Chen and Sun}]{demir2025diffdenoise}
\bibinfo{author}{Demir, B.}, \bibinfo{author}{Liu, Y.}, \bibinfo{author}{Chen, X.}, \bibinfo{author}{Chen, E.Z.}, \bibinfo{author}{Zhao, L.}, \bibinfo{author}{Mailhe, B.}, \bibinfo{author}{Chen, T.}, \bibinfo{author}{Sun, S.}, \bibinfo{year}{2025}.
\newblock \bibinfo{title}{Diffdenoise: self-supervised medical image denoising with conditional diffusion models}.
\newblock \bibinfo{journal}{arXiv preprint arXiv:2504.00264} .
\bibitem[{Frost et~al.(1982)Frost, Stiles, Shanmugan and Holtzman}]{frost1982model}
\bibinfo{author}{Frost, V.S.}, \bibinfo{author}{Stiles, J.A.}, \bibinfo{author}{Shanmugan, K.S.}, \bibinfo{author}{Holtzman, J.C.}, \bibinfo{year}{1982}.
\newblock \bibinfo{title}{A model for radar images and its application to adaptive digital filtering of multiplicative noise}.
\newblock \bibinfo{journal}{IEEE Transactions on pattern analysis and machine intelligence} , \bibinfo{pages}{157--166}.
\bibitem[{Garcia(2022)}]{garcia2022simus}
\bibinfo{author}{Garcia, D.}, \bibinfo{year}{2022}.
\newblock \bibinfo{title}{Simus: an open-source simulator for medical ultrasound imaging. part i: theory \& examples}.
\newblock \bibinfo{journal}{Computer Methods and Programs in Biomedicine} \bibinfo{volume}{218}, \bibinfo{pages}{106726}.
\bibitem[{Goudarzi and Rivaz(2023)}]{goudarzi2023deep}
\bibinfo{author}{Goudarzi, S.}, \bibinfo{author}{Rivaz, H.}, \bibinfo{year}{2023}.
\newblock \bibinfo{title}{Deep ultrasound denoising without clean data}, in: \bibinfo{booktitle}{Medical Imaging 2023: Ultrasonic Imaging and Tomography}, \bibinfo{organization}{SPIE}. pp. \bibinfo{pages}{131--136}.
\bibitem[{Gu et~al.(2019)Gu, Li, Gool and Timofte}]{gu2019self}
\bibinfo{author}{Gu, S.}, \bibinfo{author}{Li, Y.}, \bibinfo{author}{Gool, L.V.}, \bibinfo{author}{Timofte, R.}, \bibinfo{year}{2019}.
\newblock \bibinfo{title}{Self-guided network for fast image denoising}, in: \bibinfo{booktitle}{Proceedings of the IEEE/CVF International Conference on Computer Vision}, pp. \bibinfo{pages}{2511--2520}.
\bibitem[{Guo et~al.(2019)Guo, Yan, Zhang, Zuo and Zhang}]{guo2019toward}
\bibinfo{author}{Guo, S.}, \bibinfo{author}{Yan, Z.}, \bibinfo{author}{Zhang, K.}, \bibinfo{author}{Zuo, W.}, \bibinfo{author}{Zhang, L.}, \bibinfo{year}{2019}.
\newblock \bibinfo{title}{Toward convolutional blind denoising of real photographs}, in: \bibinfo{booktitle}{Proceedings of the IEEE/CVF conference on computer vision and pattern recognition}, pp. \bibinfo{pages}{1712--1722}.
\bibitem[{Haralick et~al.(1973)Haralick, Shanmugam and Dinstein}]{glcm}
\bibinfo{author}{Haralick, R.M.}, \bibinfo{author}{Shanmugam, K.}, \bibinfo{author}{Dinstein, I.}, \bibinfo{year}{1973}.
\newblock \bibinfo{title}{Textural features for image classification}.
\newblock \bibinfo{journal}{IEEE Transactions on Systems, Man, and Cybernetics} \bibinfo{volume}{SMC-3}, \bibinfo{pages}{610--621}.
\newblock \DOIprefix\doi{10.1109/TSMC.1973.4309314}.
\bibitem[{He et~al.(2016)He, Zhang, Ren and Sun}]{he2016deep}
\bibinfo{author}{He, K.}, \bibinfo{author}{Zhang, X.}, \bibinfo{author}{Ren, S.}, \bibinfo{author}{Sun, J.}, \bibinfo{year}{2016}.
\newblock \bibinfo{title}{Deep residual learning for image recognition}, in: \bibinfo{booktitle}{Proceedings of the IEEE conference on computer vision and pattern recognition}, pp. \bibinfo{pages}{770--778}.
\bibitem[{Hoskins et~al.(2019)Hoskins, Martin and Thrush}]{hoskins2019diagnostic}
\bibinfo{author}{Hoskins, P.R.}, \bibinfo{author}{Martin, K.}, \bibinfo{author}{Thrush, A.}, \bibinfo{year}{2019}.
\newblock \bibinfo{title}{Diagnostic ultrasound: physics and equipment}.
\newblock \bibinfo{publisher}{CRC Press}.
\bibitem[{Huang et~al.(2023)Huang, Bi, Navab and Jiang}]{huang2023motion}
\bibinfo{author}{Huang, D.}, \bibinfo{author}{Bi, Y.}, \bibinfo{author}{Navab, N.}, \bibinfo{author}{Jiang, Z.}, \bibinfo{year}{2023}.
\newblock \bibinfo{title}{Motion magnification in robotic sonography: Enabling pulsation-aware artery segmentation}, in: \bibinfo{booktitle}{2023 IEEE/RSJ International Conference on Intelligent Robots and Systems (IROS)}, \bibinfo{organization}{IEEE}. pp. \bibinfo{pages}{6565--6570}.
\bibitem[{Huang et~al.(2025)Huang, Li, Karlas, Chu, Au, Navab and Jiang}]{huang2025vibnet}
\bibinfo{author}{Huang, D.}, \bibinfo{author}{Li, C.}, \bibinfo{author}{Karlas, A.}, \bibinfo{author}{Chu, X.}, \bibinfo{author}{Au, K.S.}, \bibinfo{author}{Navab, N.}, \bibinfo{author}{Jiang, Z.}, \bibinfo{year}{2025}.
\newblock \bibinfo{title}{Vibnet: Vibration-boosted needle detection in ultrasound images}.
\newblock \bibinfo{journal}{IEEE Transactions on Medical Imaging} .
\bibitem[{Huang et~al.(2018)Huang, Zhang and Li}]{huang2018machine}
\bibinfo{author}{Huang, Q.}, \bibinfo{author}{Zhang, F.}, \bibinfo{author}{Li, X.}, \bibinfo{year}{2018}.
\newblock \bibinfo{title}{Machine learning in ultrasound computer-aided diagnostic systems: a survey}.
\newblock \bibinfo{journal}{BioMed research international} \bibinfo{volume}{2018}, \bibinfo{pages}{5137904}.
\bibitem[{Huang et~al.(2021)Huang, Li, Jia, Lu and Liu}]{huang2021neighbor2neighbor}
\bibinfo{author}{Huang, T.}, \bibinfo{author}{Li, S.}, \bibinfo{author}{Jia, X.}, \bibinfo{author}{Lu, H.}, \bibinfo{author}{Liu, J.}, \bibinfo{year}{2021}.
\newblock \bibinfo{title}{Neighbor2neighbor: Self-supervised denoising from single noisy images}, in: \bibinfo{booktitle}{Proceedings of the IEEE/CVF conference on computer vision and pattern recognition}, pp. \bibinfo{pages}{14781--14790}.
\bibitem[{Jang et~al.(2023)Jang, Lee, Park, Kim and Cho}]{jang2023self}
\bibinfo{author}{Jang, Y.I.}, \bibinfo{author}{Lee, K.}, \bibinfo{author}{Park, G.Y.}, \bibinfo{author}{Kim, S.}, \bibinfo{author}{Cho, N.I.}, \bibinfo{year}{2023}.
\newblock \bibinfo{title}{Self-supervised image denoising with downsampled invariance loss and conditional blind-spot network}, in: \bibinfo{booktitle}{Proceedings of the IEEE/CVF International Conference on Computer Vision}, pp. \bibinfo{pages}{12196--12205}.
\bibitem[{Jiang et~al.(2024a)Jiang, Bi, Zhou, Hu, Burke and Navab}]{jiang2024intelligent}
\bibinfo{author}{Jiang, Z.}, \bibinfo{author}{Bi, Y.}, \bibinfo{author}{Zhou, M.}, \bibinfo{author}{Hu, Y.}, \bibinfo{author}{Burke, M.}, \bibinfo{author}{Navab, N.}, \bibinfo{year}{2024}a.
\newblock \bibinfo{title}{Intelligent robotic sonographer: Mutual information-based disentangled reward learning from few demonstrations}.
\newblock \bibinfo{journal}{The International Journal of Robotics Research} \bibinfo{volume}{43}, \bibinfo{pages}{981--1002}.
\bibitem[{Jiang et~al.(2024b)Jiang, Kang, Bi, Li, Li and Navab}]{jiang2024class}
\bibinfo{author}{Jiang, Z.}, \bibinfo{author}{Kang, Y.}, \bibinfo{author}{Bi, Y.}, \bibinfo{author}{Li, X.}, \bibinfo{author}{Li, C.}, \bibinfo{author}{Navab, N.}, \bibinfo{year}{2024}b.
\newblock \bibinfo{title}{Class-aware cartilage segmentation for autonomous us-ct registration in robotic intercostal ultrasound imaging}.
\newblock \bibinfo{journal}{IEEE Transactions on Automation Science and Engineering} .
\bibitem[{Jiang et~al.(2023a)Jiang, Salcudean and Navab}]{jiang2023robotic}
\bibinfo{author}{Jiang, Z.}, \bibinfo{author}{Salcudean, S.E.}, \bibinfo{author}{Navab, N.}, \bibinfo{year}{2023}a.
\newblock \bibinfo{title}{Robotic ultrasound imaging: State-of-the-art and future perspectives}.
\newblock \bibinfo{journal}{Medical Image Analysis} , \bibinfo{pages}{102878}.
\bibitem[{Jiang et~al.(2021)Jiang, Zhou, Bi, Zhou, Wendler and Navab}]{jiang2021deformation}
\bibinfo{author}{Jiang, Z.}, \bibinfo{author}{Zhou, Y.}, \bibinfo{author}{Bi, Y.}, \bibinfo{author}{Zhou, M.}, \bibinfo{author}{Wendler, T.}, \bibinfo{author}{Navab, N.}, \bibinfo{year}{2021}.
\newblock \bibinfo{title}{Deformation-aware robotic 3d ultrasound}.
\newblock \bibinfo{journal}{IEEE Robotics and Automation Letters} \bibinfo{volume}{6}, \bibinfo{pages}{7675--7682}.
\bibitem[{Jiang et~al.(2023b)Jiang, Zhou, Cao and Navab}]{jiang2023defcor}
\bibinfo{author}{Jiang, Z.}, \bibinfo{author}{Zhou, Y.}, \bibinfo{author}{Cao, D.}, \bibinfo{author}{Navab, N.}, \bibinfo{year}{2023}b.
\newblock \bibinfo{title}{Defcor-net: Physics-aware ultrasound deformation correction}.
\newblock \bibinfo{journal}{Medical Image Analysis} \bibinfo{volume}{90}, \bibinfo{pages}{102923}.
\bibitem[{Jung et~al.(2024)Jung, Kang, Park, Guezzi and Yu}]{jung2024unsupervised}
\bibinfo{author}{Jung, D.}, \bibinfo{author}{Kang, M.}, \bibinfo{author}{Park, S.H.}, \bibinfo{author}{Guezzi, N.}, \bibinfo{author}{Yu, J.}, \bibinfo{year}{2024}.
\newblock \bibinfo{title}{Unsupervised speckle noise reduction technique for clinical ultrasound imaging}.
\newblock \bibinfo{journal}{Ultrasonography} \bibinfo{volume}{43}, \bibinfo{pages}{327}.
\bibitem[{Kang et~al.(2024)Kang, Lao, Gao, Liu, Yi, Ma, Zhang and Li}]{kang2024deblurring}
\bibinfo{author}{Kang, Q.}, \bibinfo{author}{Lao, Q.}, \bibinfo{author}{Gao, J.}, \bibinfo{author}{Liu, J.}, \bibinfo{author}{Yi, H.}, \bibinfo{author}{Ma, B.}, \bibinfo{author}{Zhang, X.}, \bibinfo{author}{Li, K.}, \bibinfo{year}{2024}.
\newblock \bibinfo{title}{Deblurring masked image modeling for ultrasound image analysis}.
\newblock \bibinfo{journal}{Medical Image Analysis} \bibinfo{volume}{97}, \bibinfo{pages}{103256}.
\bibitem[{Krissian et~al.(2005)Krissian, Kikinis, Westin and Vosburgh}]{krissian2005speckle}
\bibinfo{author}{Krissian, K.}, \bibinfo{author}{Kikinis, R.}, \bibinfo{author}{Westin, C.F.}, \bibinfo{author}{Vosburgh, K.}, \bibinfo{year}{2005}.
\newblock \bibinfo{title}{Speckle-constrained filtering of ultrasound images}, in: \bibinfo{booktitle}{2005 IEEE Computer Society Conference on Computer Vision and Pattern Recognition (CVPR'05)}, \bibinfo{organization}{IEEE}. pp. \bibinfo{pages}{547--552}.
\bibitem[{Krissian et~al.(2007)Krissian, Westin, Kikinis and Vosburgh}]{krissian2007oriented}
\bibinfo{author}{Krissian, K.}, \bibinfo{author}{Westin, C.F.}, \bibinfo{author}{Kikinis, R.}, \bibinfo{author}{Vosburgh, K.G.}, \bibinfo{year}{2007}.
\newblock \bibinfo{title}{Oriented speckle reducing anisotropic diffusion}.
\newblock \bibinfo{journal}{IEEE Transactions on Image Processing} \bibinfo{volume}{16}, \bibinfo{pages}{1412--1424}.
\bibitem[{Krull et~al.(2019)Krull, Buchholz and Jug}]{krull2019noise2void}
\bibinfo{author}{Krull, A.}, \bibinfo{author}{Buchholz, T.O.}, \bibinfo{author}{Jug, F.}, \bibinfo{year}{2019}.
\newblock \bibinfo{title}{Noise2void-learning denoising from single noisy images}, in: \bibinfo{booktitle}{Proceedings of the IEEE/CVF conference on computer vision and pattern recognition}, pp. \bibinfo{pages}{2129--2137}.
\bibitem[{Laine et~al.(2019)Laine, Karras, Lehtinen and Aila}]{laine2019high}
\bibinfo{author}{Laine, S.}, \bibinfo{author}{Karras, T.}, \bibinfo{author}{Lehtinen, J.}, \bibinfo{author}{Aila, T.}, \bibinfo{year}{2019}.
\newblock \bibinfo{title}{High-quality self-supervised deep image denoising}.
\newblock \bibinfo{journal}{Advances in Neural Information Processing Systems} \bibinfo{volume}{32}.
\bibitem[{Lee et~al.(2022a)Lee, Lee, Youn, Lee, Lew and Hwang}]{lee2022speckle}
\bibinfo{author}{Lee, H.}, \bibinfo{author}{Lee, M.H.}, \bibinfo{author}{Youn, S.}, \bibinfo{author}{Lee, K.}, \bibinfo{author}{Lew, H.M.}, \bibinfo{author}{Hwang, J.Y.}, \bibinfo{year}{2022}a.
\newblock \bibinfo{title}{Speckle reduction via deep content-aware image prior for precise breast tumor segmentation in an ultrasound image}.
\newblock \bibinfo{journal}{IEEE Transactions on Ultrasonics, Ferroelectrics, and Frequency Control} \bibinfo{volume}{69}, \bibinfo{pages}{2638--2650}.
\bibitem[{Lee et~al.(2022b)Lee, Son and Lee}]{lee2022ap}
\bibinfo{author}{Lee, W.}, \bibinfo{author}{Son, S.}, \bibinfo{author}{Lee, K.M.}, \bibinfo{year}{2022}b.
\newblock \bibinfo{title}{Ap-bsn: Self-supervised denoising for real-world images via asymmetric pd and blind-spot network}, in: \bibinfo{booktitle}{Proceedings of the IEEE/CVF Conference on Computer Vision and Pattern Recognition}, pp. \bibinfo{pages}{17725--17734}.
\bibitem[{Lefkimmiatis(2018)}]{lefkimmiatis2018universal}
\bibinfo{author}{Lefkimmiatis, S.}, \bibinfo{year}{2018}.
\newblock \bibinfo{title}{Universal denoising networks: a novel cnn architecture for image denoising}, in: \bibinfo{booktitle}{Proceedings of the IEEE conference on computer vision and pattern recognition}, pp. \bibinfo{pages}{3204--3213}.
\bibitem[{Lefkimmiatis and Koshelev(2023)}]{lefkimmiatis2023learning}
\bibinfo{author}{Lefkimmiatis, S.}, \bibinfo{author}{Koshelev, I.}, \bibinfo{year}{2023}.
\newblock \bibinfo{title}{Learning sparse and low-rank priors for image recovery via iterative reweighted least squares minimization}.
\newblock \bibinfo{journal}{arXiv preprint arXiv:2304.10536} .
\bibitem[{Lehtinen(2018)}]{lehtinen2018noise2noise}
\bibinfo{author}{Lehtinen, J.}, \bibinfo{year}{2018}.
\newblock \bibinfo{title}{Noise2noise: Learning image restoration without clean data}.
\newblock \bibinfo{journal}{arXiv preprint arXiv:1803.04189} .
\bibitem[{Li et~al.(2024)Li, Su, Zhao, Lv, Wang, Navab, Hu and Jiang}]{li2024ultrasound}
\bibinfo{author}{Li, J.}, \bibinfo{author}{Su, T.}, \bibinfo{author}{Zhao, B.}, \bibinfo{author}{Lv, F.}, \bibinfo{author}{Wang, Q.}, \bibinfo{author}{Navab, N.}, \bibinfo{author}{Hu, Y.}, \bibinfo{author}{Jiang, Z.}, \bibinfo{year}{2024}.
\newblock \bibinfo{title}{Ultrasound report generation with cross-modality feature alignment via unsupervised guidance}.
\newblock \bibinfo{journal}{IEEE Transactions on Medical Imaging} .
\bibitem[{Li et~al.(2025)Li, Huang, Zhang, Navab and Jiang}]{li2025semantic}
\bibinfo{author}{Li, X.}, \bibinfo{author}{Huang, D.}, \bibinfo{author}{Zhang, Y.}, \bibinfo{author}{Navab, N.}, \bibinfo{author}{Jiang, Z.}, \bibinfo{year}{2025}.
\newblock \bibinfo{title}{Semantic scene graph for ultrasound image explanation and scanning guidance}, in: \bibinfo{booktitle}{International Conference on Medical Image Computing and Computer-Assisted Intervention}, \bibinfo{organization}{Springer}.
\bibitem[{Lu and et~al.(2016)}]{lu2016tensor}
\bibinfo{author}{Lu, C.}, \bibinfo{author}{et~al.}, \bibinfo{year}{2016}.
\newblock \bibinfo{title}{Tensor robust principal component analysis: Exact recovery of corrupted low-rank tensors via convex optimization}, in: \bibinfo{booktitle}{CVPR}, pp. \bibinfo{pages}{5249--5257}.
\bibitem[{Lu et~al.(2023)Lu, Millioz, Varray, Por{\'e}e, Provost, Bernard, Garcia and Friboulet}]{lu2023ultrafast}
\bibinfo{author}{Lu, J.}, \bibinfo{author}{Millioz, F.}, \bibinfo{author}{Varray, F.}, \bibinfo{author}{Por{\'e}e, J.}, \bibinfo{author}{Provost, J.}, \bibinfo{author}{Bernard, O.}, \bibinfo{author}{Garcia, D.}, \bibinfo{author}{Friboulet, D.}, \bibinfo{year}{2023}.
\newblock \bibinfo{title}{Ultrafast cardiac imaging using deep learning for speckle-tracking echocardiography}.
\newblock \bibinfo{journal}{IEEE Transactions on Ultrasonics, Ferroelectrics, and Frequency Control} \bibinfo{volume}{70}, \bibinfo{pages}{1761--1772}.
\bibitem[{Mansour and Heckel(2023)}]{mansour2023zero}
\bibinfo{author}{Mansour, Y.}, \bibinfo{author}{Heckel, R.}, \bibinfo{year}{2023}.
\newblock \bibinfo{title}{Zero-shot noise2noise: Efficient image denoising without any data}, in: \bibinfo{booktitle}{Proceedings of the IEEE/CVF Conference on Computer Vision and Pattern Recognition}, pp. \bibinfo{pages}{14018--14027}.
\bibitem[{Mao et~al.(2016)Mao, Shen and Yang}]{mao2016image}
\bibinfo{author}{Mao, X.}, \bibinfo{author}{Shen, C.}, \bibinfo{author}{Yang, Y.B.}, \bibinfo{year}{2016}.
\newblock \bibinfo{title}{Image restoration using very deep convolutional encoder-decoder networks with symmetric skip connections}.
\newblock \bibinfo{journal}{Advances in neural information processing systems} \bibinfo{volume}{29}.
\bibitem[{Martin et~al.(2001)Martin, Fowlkes, Tal and Malik}]{MartinFTM01}
\bibinfo{author}{Martin, D.}, \bibinfo{author}{Fowlkes, C.}, \bibinfo{author}{Tal, D.}, \bibinfo{author}{Malik, J.}, \bibinfo{year}{2001}.
\newblock \bibinfo{title}{A database of human segmented natural images and its application to evaluating segmentation algorithms and measuring ecological statistics}, in: \bibinfo{booktitle}{Proc. 8th Int'l Conf. Computer Vision}, pp. \bibinfo{pages}{416--423}.
\bibitem[{Mei et~al.(2019)Mei, Hu, Fei and Qin}]{mei2019phase}
\bibinfo{author}{Mei, K.}, \bibinfo{author}{Hu, B.}, \bibinfo{author}{Fei, B.}, \bibinfo{author}{Qin, B.}, \bibinfo{year}{2019}.
\newblock \bibinfo{title}{Phase asymmetry ultrasound despeckling with fractional anisotropic diffusion and total variation}.
\newblock \bibinfo{journal}{IEEE Transactions on Image Processing} \bibinfo{volume}{29}, \bibinfo{pages}{2845--2859}.
\bibitem[{Mwikirize et~al.(2018)Mwikirize, Nosher and Hacihaliloglu}]{mwikirize2018signal}
\bibinfo{author}{Mwikirize, C.}, \bibinfo{author}{Nosher, J.L.}, \bibinfo{author}{Hacihaliloglu, I.}, \bibinfo{year}{2018}.
\newblock \bibinfo{title}{Signal attenuation maps for needle enhancement and localization in 2d ultrasound}.
\newblock \bibinfo{journal}{International journal of computer assisted radiology and surgery} \bibinfo{volume}{13}, \bibinfo{pages}{363--374}.
\bibitem[{Nagare et~al.(2021)Nagare, Melnyk, Rahman, Sauer and Bouman}]{nagare2021bias}
\bibinfo{author}{Nagare, M.}, \bibinfo{author}{Melnyk, R.}, \bibinfo{author}{Rahman, O.}, \bibinfo{author}{Sauer, K.D.}, \bibinfo{author}{Bouman, C.A.}, \bibinfo{year}{2021}.
\newblock \bibinfo{title}{A bias-reducing loss function for ct image denoising}, in: \bibinfo{booktitle}{ICASSP 2021-2021 IEEE International Conference on Acoustics, Speech and Signal Processing (ICASSP)}, \bibinfo{organization}{IEEE}. pp. \bibinfo{pages}{1175--1179}.
\bibitem[{Nyrnes et~al.(2020)Nyrnes, Fadnes, Wigen, Mertens and Lovstakken}]{nyrnes2020blood}
\bibinfo{author}{Nyrnes, S.A.}, \bibinfo{author}{Fadnes, S.}, \bibinfo{author}{Wigen, M.S.}, \bibinfo{author}{Mertens, L.}, \bibinfo{author}{Lovstakken, L.}, \bibinfo{year}{2020}.
\newblock \bibinfo{title}{Blood speckle-tracking based on high--frame rate ultrasound imaging in pediatric cardiology}.
\newblock \bibinfo{journal}{Journal of the American Society of Echocardiography} \bibinfo{volume}{33}, \bibinfo{pages}{493--503}.
\bibitem[{Perona and Malik(1990)}]{perona1990scale}
\bibinfo{author}{Perona, P.}, \bibinfo{author}{Malik, J.}, \bibinfo{year}{1990}.
\newblock \bibinfo{title}{Scale-space and edge detection using anisotropic diffusion}.
\newblock \bibinfo{journal}{IEEE Transactions on pattern analysis and machine intelligence} \bibinfo{volume}{12}, \bibinfo{pages}{629--639}.
\bibitem[{Perrot et~al.(2021)Perrot, Polichetti, Varray and Garcia}]{perrot2021so}
\bibinfo{author}{Perrot, V.}, \bibinfo{author}{Polichetti, M.}, \bibinfo{author}{Varray, F.}, \bibinfo{author}{Garcia, D.}, \bibinfo{year}{2021}.
\newblock \bibinfo{title}{So you think you can das? a viewpoint on delay-and-sum beamforming}.
\newblock \bibinfo{journal}{Ultrasonics} \bibinfo{volume}{111}, \bibinfo{pages}{106309}.
\bibitem[{Pl{\"o}tz and Roth(2018)}]{plotz2018neural}
\bibinfo{author}{Pl{\"o}tz, T.}, \bibinfo{author}{Roth, S.}, \bibinfo{year}{2018}.
\newblock \bibinfo{title}{Neural nearest neighbors networks}.
\newblock \bibinfo{journal}{Advances in Neural information processing systems} \bibinfo{volume}{31}.
\bibitem[{Prince and Links(2006)}]{prince2006medical}
\bibinfo{author}{Prince, J.L.}, \bibinfo{author}{Links, J.M.}, \bibinfo{year}{2006}.
\newblock \bibinfo{title}{Medical imaging signals and systems}. volume~\bibinfo{volume}{37}.
\newblock \bibinfo{publisher}{Pearson Prentice Hall Upper Saddle River}.
\bibitem[{Ramalhinho et~al.(2020)Ramalhinho, Tregidgo, Gurusamy, Hawkes, Davidson and Clarkson}]{ramalhinho2020registration}
\bibinfo{author}{Ramalhinho, J.}, \bibinfo{author}{Tregidgo, H.F.}, \bibinfo{author}{Gurusamy, K.}, \bibinfo{author}{Hawkes, D.J.}, \bibinfo{author}{Davidson, B.}, \bibinfo{author}{Clarkson, M.J.}, \bibinfo{year}{2020}.
\newblock \bibinfo{title}{Registration of untracked 2d laparoscopic ultrasound to ct images of the liver using multi-labelled content-based image retrieval}.
\newblock \bibinfo{journal}{IEEE Transactions on Medical Imaging} \bibinfo{volume}{40}, \bibinfo{pages}{1042--1054}.
\bibitem[{Ravi et~al.(2024)Ravi, Gabeur, Hu, Hu, Ryali, Ma, Khedr, R{\"a}dle, Rolland, Gustafson et~al.}]{ravi2024sam}
\bibinfo{author}{Ravi, N.}, \bibinfo{author}{Gabeur, V.}, \bibinfo{author}{Hu, Y.T.}, \bibinfo{author}{Hu, R.}, \bibinfo{author}{Ryali, C.}, \bibinfo{author}{Ma, T.}, \bibinfo{author}{Khedr, H.}, \bibinfo{author}{R{\"a}dle, R.}, \bibinfo{author}{Rolland, C.}, \bibinfo{author}{Gustafson, L.}, et~al., \bibinfo{year}{2024}.
\newblock \bibinfo{title}{Sam 2: Segment anything in images and videos}.
\newblock \bibinfo{journal}{arXiv preprint arXiv:2408.00714} .
\bibitem[{Ronneberger et~al.(2015)Ronneberger, Fischer and Brox}]{ronneberger2015u}
\bibinfo{author}{Ronneberger, O.}, \bibinfo{author}{Fischer, P.}, \bibinfo{author}{Brox, T.}, \bibinfo{year}{2015}.
\newblock \bibinfo{title}{U-net: Convolutional networks for biomedical image segmentation}, in: \bibinfo{booktitle}{International Conference on Medical image computing and computer-assisted intervention}, \bibinfo{organization}{Springer}. pp. \bibinfo{pages}{234--241}.
\bibitem[{Sagheer and George(2017)}]{sagheer2017ultrasound}
\bibinfo{author}{Sagheer, S.V.M.}, \bibinfo{author}{George, S.N.}, \bibinfo{year}{2017}.
\newblock \bibinfo{title}{Ultrasound image despeckling using low rank matrix approximation approach}.
\newblock \bibinfo{journal}{Biomedical Signal Processing and Control} \bibinfo{volume}{38}, \bibinfo{pages}{236--249}.
\bibitem[{Sharifzadeh et~al.(2024)Sharifzadeh, Goudarzi, Tang, Benali and Rivaz}]{sharifzadeh2024mitigating}
\bibinfo{author}{Sharifzadeh, M.}, \bibinfo{author}{Goudarzi, S.}, \bibinfo{author}{Tang, A.}, \bibinfo{author}{Benali, H.}, \bibinfo{author}{Rivaz, H.}, \bibinfo{year}{2024}.
\newblock \bibinfo{title}{Mitigating aberration-induced noise: A deep learning-based aberration-to-aberration approach}.
\newblock \bibinfo{journal}{IEEE Transactions on Medical Imaging} .
\bibitem[{Song et~al.(2021)Song, Guo, Xu, Chao, Xu, Turkbey, Wood, Wang and Yan}]{song2021cross}
\bibinfo{author}{Song, X.}, \bibinfo{author}{Guo, H.}, \bibinfo{author}{Xu, X.}, \bibinfo{author}{Chao, H.}, \bibinfo{author}{Xu, S.}, \bibinfo{author}{Turkbey, B.}, \bibinfo{author}{Wood, B.J.}, \bibinfo{author}{Wang, G.}, \bibinfo{author}{Yan, P.}, \bibinfo{year}{2021}.
\newblock \bibinfo{title}{Cross-modal attention for mri and ultrasound volume registration}, in: \bibinfo{booktitle}{Medical Image Computing and Computer Assisted Intervention--MICCAI 2021: 24th International Conference, Strasbourg, France, September 27--October 1, 2021, Proceedings, Part IV 24}, \bibinfo{organization}{Springer}. pp. \bibinfo{pages}{66--75}.
\bibitem[{Stevens et~al.(2024)Stevens, Meral, Yu, Apostolakis, Robert and Van~Sloun}]{stevens2024dehazing}
\bibinfo{author}{Stevens, T.S.}, \bibinfo{author}{Meral, F.C.}, \bibinfo{author}{Yu, J.}, \bibinfo{author}{Apostolakis, I.Z.}, \bibinfo{author}{Robert, J.L.}, \bibinfo{author}{Van~Sloun, R.J.}, \bibinfo{year}{2024}.
\newblock \bibinfo{title}{Dehazing ultrasound using diffusion models}.
\newblock \bibinfo{journal}{IEEE Transactions on Medical Imaging} \bibinfo{volume}{43}, \bibinfo{pages}{3546--3558}.
\bibitem[{Szabo(2013)}]{szabo2013diagnostic}
\bibinfo{author}{Szabo, T.L.}, \bibinfo{year}{2013}.
\newblock \bibinfo{title}{Diagnostic ultrasound imaging: inside out}.
\newblock \bibinfo{publisher}{Academic press}.
\bibitem[{Tai et~al.(2017)Tai, Yang, Liu and Xu}]{tai2017memnet}
\bibinfo{author}{Tai, Y.}, \bibinfo{author}{Yang, J.}, \bibinfo{author}{Liu, X.}, \bibinfo{author}{Xu, C.}, \bibinfo{year}{2017}.
\newblock \bibinfo{title}{Memnet: A persistent memory network for image restoration}, in: \bibinfo{booktitle}{Proceedings of the IEEE international conference on computer vision}, pp. \bibinfo{pages}{4539--4547}.
\bibitem[{Tay et~al.(2010)Tay, Garson, Acton and Hossack}]{tay2010ultrasound}
\bibinfo{author}{Tay, P.C.}, \bibinfo{author}{Garson, C.D.}, \bibinfo{author}{Acton, S.T.}, \bibinfo{author}{Hossack, J.A.}, \bibinfo{year}{2010}.
\newblock \bibinfo{title}{Ultrasound despeckling for contrast enhancement}.
\newblock \bibinfo{journal}{IEEE Transactions on Image Processing} \bibinfo{volume}{19}, \bibinfo{pages}{1847--1860}.
\bibitem[{Ulyanov et~al.(2018)Ulyanov, Vedaldi and Lempitsky}]{ulyanov2018deep}
\bibinfo{author}{Ulyanov, D.}, \bibinfo{author}{Vedaldi, A.}, \bibinfo{author}{Lempitsky, V.}, \bibinfo{year}{2018}.
\newblock \bibinfo{title}{Deep image prior}, in: \bibinfo{booktitle}{Proceedings of the IEEE conference on computer vision and pattern recognition}, pp. \bibinfo{pages}{9446--9454}.
\bibitem[{Wagner et~al.(1983)Wagner, Smith, Sandrik and Lopez}]{wagner1983statistics}
\bibinfo{author}{Wagner, R.F.}, \bibinfo{author}{Smith, S.W.}, \bibinfo{author}{Sandrik, J.M.}, \bibinfo{author}{Lopez, H.}, \bibinfo{year}{1983}.
\newblock \bibinfo{title}{Statistics of speckle in ultrasound b-scans}.
\newblock \bibinfo{journal}{IEEE Transactions on sonics and ultrasonics} \bibinfo{volume}{30}, \bibinfo{pages}{156--163}.
\bibitem[{Wen et~al.(2023)Wen, Peng, Wei, Luo and Jiang}]{wen2023convolutional}
\bibinfo{author}{Wen, S.}, \bibinfo{author}{Peng, B.}, \bibinfo{author}{Wei, X.}, \bibinfo{author}{Luo, J.}, \bibinfo{author}{Jiang, J.}, \bibinfo{year}{2023}.
\newblock \bibinfo{title}{Convolutional neural network-based speckle tracking for ultrasound strain elastography: An unsupervised learning approach}.
\newblock \bibinfo{journal}{IEEE Transactions on Ultrasonics, Ferroelectrics, and Frequency Control} \bibinfo{volume}{70}, \bibinfo{pages}{354--367}.
\bibitem[{Wright et~al.(2009)Wright, Ganesh, Rao, Peng and Ma}]{wright2009robust}
\bibinfo{author}{Wright, J.}, \bibinfo{author}{Ganesh, A.}, \bibinfo{author}{Rao, S.}, \bibinfo{author}{Peng, Y.}, \bibinfo{author}{Ma, Y.}, \bibinfo{year}{2009}.
\newblock \bibinfo{title}{Robust principal component analysis: Exact recovery of corrupted low-rank matrices via convex optimization}.
\newblock \bibinfo{journal}{Advances in neural information processing systems} \bibinfo{volume}{22}.
\bibitem[{Wu et~al.(2020)Wu, Liu, Cao, Ren and Zuo}]{wu2020unpaired}
\bibinfo{author}{Wu, X.}, \bibinfo{author}{Liu, M.}, \bibinfo{author}{Cao, Y.}, \bibinfo{author}{Ren, D.}, \bibinfo{author}{Zuo, W.}, \bibinfo{year}{2020}.
\newblock \bibinfo{title}{Unpaired learning of deep image denoising}, in: \bibinfo{booktitle}{European conference on computer vision}, \bibinfo{organization}{Springer}. pp. \bibinfo{pages}{352--368}.
\bibitem[{Xie et~al.(2016)Xie, Zhao, Meng, Xu, Gu, Zuo and Zhang}]{xie2016multispectral}
\bibinfo{author}{Xie, Q.}, \bibinfo{author}{Zhao, Q.}, \bibinfo{author}{Meng, D.}, \bibinfo{author}{Xu, Z.}, \bibinfo{author}{Gu, S.}, \bibinfo{author}{Zuo, W.}, \bibinfo{author}{Zhang, L.}, \bibinfo{year}{2016}.
\newblock \bibinfo{title}{Multispectral images denoising by intrinsic tensor sparsity regularization}, in: \bibinfo{booktitle}{Proceedings of the IEEE conference on computer vision and pattern recognition}, pp. \bibinfo{pages}{1692--1700}.
\bibitem[{Yang et~al.(2022)Yang, Guo, Chen and Yuan}]{yang2022source}
\bibinfo{author}{Yang, C.}, \bibinfo{author}{Guo, X.}, \bibinfo{author}{Chen, Z.}, \bibinfo{author}{Yuan, Y.}, \bibinfo{year}{2022}.
\newblock \bibinfo{title}{Source free domain adaptation for medical image segmentation with fourier style mining}.
\newblock \bibinfo{journal}{Medical Image Analysis} \bibinfo{volume}{79}, \bibinfo{pages}{102457}.
\bibitem[{Yang et~al.(2024)Yang, Li, Xu, Tang, Wang, Tsui and Zhou}]{yang2024frequency}
\bibinfo{author}{Yang, K.}, \bibinfo{author}{Li, Q.}, \bibinfo{author}{Xu, J.}, \bibinfo{author}{Tang, M.X.}, \bibinfo{author}{Wang, Z.}, \bibinfo{author}{Tsui, P.H.}, \bibinfo{author}{Zhou, X.}, \bibinfo{year}{2024}.
\newblock \bibinfo{title}{Frequency-domain robust pca for real-time monitoring of hifu treatment}.
\newblock \bibinfo{journal}{IEEE Transactions on Medical Imaging} .
\bibitem[{Yu and Acton(2002)}]{yu2002speckle}
\bibinfo{author}{Yu, Y.}, \bibinfo{author}{Acton, S.T.}, \bibinfo{year}{2002}.
\newblock \bibinfo{title}{Speckle reducing anisotropic diffusion}.
\newblock \bibinfo{journal}{IEEE Transactions on image processing} \bibinfo{volume}{11}, \bibinfo{pages}{1260--1270}.
\bibitem[{Zhang et~al.(2015)Zhang, Lin, Wu, Wang and Cheng}]{zhang2015wavelet}
\bibinfo{author}{Zhang, J.}, \bibinfo{author}{Lin, G.}, \bibinfo{author}{Wu, L.}, \bibinfo{author}{Wang, C.}, \bibinfo{author}{Cheng, Y.}, \bibinfo{year}{2015}.
\newblock \bibinfo{title}{Wavelet and fast bilateral filter based de-speckling method for medical ultrasound images}.
\newblock \bibinfo{journal}{Biomedical Signal Processing and Control} \bibinfo{volume}{18}, \bibinfo{pages}{1--10}.
\bibitem[{Zhang et~al.(2017)Zhang, Zuo, Chen, Meng and Zhang}]{zhang2017beyond}
\bibinfo{author}{Zhang, K.}, \bibinfo{author}{Zuo, W.}, \bibinfo{author}{Chen, Y.}, \bibinfo{author}{Meng, D.}, \bibinfo{author}{Zhang, L.}, \bibinfo{year}{2017}.
\newblock \bibinfo{title}{Beyond a gaussian denoiser: Residual learning of deep cnn for image denoising}.
\newblock \bibinfo{journal}{IEEE transactions on image processing} \bibinfo{volume}{26}, \bibinfo{pages}{3142--3155}.
\bibitem[{Zhang et~al.(2018)Zhang, Isola, Efros, Shechtman and Wang}]{zhang2018unreasonable}
\bibinfo{author}{Zhang, R.}, \bibinfo{author}{Isola, P.}, \bibinfo{author}{Efros, A.A.}, \bibinfo{author}{Shechtman, E.}, \bibinfo{author}{Wang, O.}, \bibinfo{year}{2018}.
\newblock \bibinfo{title}{The unreasonable effectiveness of deep features as a perceptual metric}, in: \bibinfo{booktitle}{Proceedings of the IEEE conference on computer vision and pattern recognition}, pp. \bibinfo{pages}{586--595}.
\bibitem[{Zheng et~al.(2012)Zheng, Liu, Sugimoto, Yan and Okutomi}]{zheng2012practical}
\bibinfo{author}{Zheng, Y.}, \bibinfo{author}{Liu, G.}, \bibinfo{author}{Sugimoto, S.}, \bibinfo{author}{Yan, S.}, \bibinfo{author}{Okutomi, M.}, \bibinfo{year}{2012}.
\newblock \bibinfo{title}{Practical low-rank matrix approximation under robust l 1-norm}, in: \bibinfo{booktitle}{2012 IEEE Conference on Computer Vision and Pattern Recognition}, \bibinfo{organization}{IEEE}. pp. \bibinfo{pages}{1410--1417}.
\bibitem[{Zhou et~al.(2025)Zhou, Bi, Tong, Wang, Navab and Jiang}]{zhou2025ultraad}
\bibinfo{author}{Zhou, Y.}, \bibinfo{author}{Bi, Y.}, \bibinfo{author}{Tong, W.}, \bibinfo{author}{Wang, W.}, \bibinfo{author}{Navab, N.}, \bibinfo{author}{Jiang, Z.}, \bibinfo{year}{2025}.
\newblock \bibinfo{title}{Ultraad: Fine-grained ultrasound anomaly classification via few-shot clip adaptation}, in: \bibinfo{booktitle}{International Conference on Medical Image Computing and Computer-Assisted Intervention}, \bibinfo{organization}{Springer}.
\bibitem[{Zhou et~al.(2019)Zhou, Zang, Xu, He, Lu and Fang}]{zhou2019iterative}
\bibinfo{author}{Zhou, Y.}, \bibinfo{author}{Zang, H.}, \bibinfo{author}{Xu, S.}, \bibinfo{author}{He, H.}, \bibinfo{author}{Lu, J.}, \bibinfo{author}{Fang, H.}, \bibinfo{year}{2019}.
\newblock \bibinfo{title}{An iterative speckle filtering algorithm for ultrasound images based on bayesian nonlocal means filter model}.
\newblock \bibinfo{journal}{Biomedical signal processing and control} \bibinfo{volume}{48}, \bibinfo{pages}{104--117}.
\bibitem[{Zhou et~al.(2014)Zhou, Guo, Dong, Sun, Zhang and Wu}]{zhou2014doubly}
\bibinfo{author}{Zhou, Z.}, \bibinfo{author}{Guo, Z.}, \bibinfo{author}{Dong, G.}, \bibinfo{author}{Sun, J.}, \bibinfo{author}{Zhang, D.}, \bibinfo{author}{Wu, B.}, \bibinfo{year}{2014}.
\newblock \bibinfo{title}{A doubly degenerate diffusion model based on the gray level indicator for multiplicative noise removal}.
\newblock \bibinfo{journal}{IEEE Transactions on Image Processing} \bibinfo{volume}{24}, \bibinfo{pages}{249--260}.
\bibitem[{Zhu et~al.(2017)Zhu, Fu, Brown and Heng}]{zhu2017non}
\bibinfo{author}{Zhu, L.}, \bibinfo{author}{Fu, C.W.}, \bibinfo{author}{Brown, M.S.}, \bibinfo{author}{Heng, P.A.}, \bibinfo{year}{2017}.
\newblock \bibinfo{title}{A non-local low-rank framework for ultrasound speckle reduction}, in: \bibinfo{booktitle}{Proceedings of the IEEE conference on computer vision and pattern recognition}, pp. \bibinfo{pages}{5650--5658}.

\end{thebibliography}



\end{document}